\documentclass[preprint,12pt]{elsarticle}

\usepackage{xcolor}
\usepackage{xspace}
\usepackage{amsfonts}
\usepackage{caption}
\usepackage{subcaption}
\usepackage{placeins}

\usepackage{url}
\usepackage{amssymb}
\usepackage{amsmath}
\usepackage{algorithm}
\usepackage{algorithmic}

\journal{}

\definecolor {snow}                {rgb}{1.00,0.98,0.98}
\definecolor {ghostwhite}          {rgb}{0.97,0.97,1.00}
\definecolor {whitesmoke}          {rgb}{0.96,0.96,0.96}
\definecolor {gainsboro}           {rgb}{0.86,0.86,0.86}
\definecolor {floralwhite}         {rgb}{1.00,0.98,0.94}
\definecolor {oldlace}             {rgb}{0.99,0.96,0.90}
\definecolor {linen}               {rgb}{0.98,0.94,0.90}
\definecolor {antiquewhite}        {rgb}{0.98,0.92,0.84}
\definecolor {papayawhip}          {rgb}{1.00,0.94,0.84}
\definecolor {blanchedalmond}      {rgb}{1.00,0.92,0.80}
\definecolor {bisque}              {rgb}{1.00,0.89,0.77}
\definecolor {peachpuff}           {rgb}{1.00,0.85,0.73}
\definecolor {navajowhite}         {rgb}{1.00,0.87,0.68}
\definecolor {moccasin}            {rgb}{1.00,0.89,0.71}
\definecolor {cornsilk}            {rgb}{1.00,0.97,0.86}
\definecolor {ivory}               {rgb}{1.00,1.00,0.94}
\definecolor {lemonchiffon}        {rgb}{1.00,0.98,0.80}
\definecolor {seashell}            {rgb}{1.00,0.96,0.93}
\definecolor {honeydew}            {rgb}{0.94,1.00,0.94}
\definecolor {mintcream}           {rgb}{0.96,1.00,0.98}
\definecolor {azure}               {rgb}{0.94,1.00,1.00}
\definecolor {aliceblue}           {rgb}{0.94,0.97,1.00}
\definecolor {lavender}            {rgb}{0.90,0.90,0.98}
\definecolor {lavenderblush}       {rgb}{1.00,0.94,0.96}
\definecolor {mistyrose}           {rgb}{1.00,0.89,0.88}
\definecolor {white}               {rgb}{1.00,1.00,1.00}
\definecolor {black}               {rgb}{0.00,0.00,0.00}
\definecolor {darkslategray}       {rgb}{0.18,0.31,0.31}
\definecolor {dimgray}             {rgb}{0.41,0.41,0.41}
\definecolor {slategray}           {rgb}{0.44,0.50,0.56}
\definecolor {lightslategray}      {rgb}{0.47,0.53,0.60}
\definecolor {gray}                {rgb}{0.75,0.75,0.75}
\definecolor {lightgrey}           {rgb}{0.83,0.83,0.83}
\definecolor {midnightblue}        {rgb}{0.10,0.10,0.44}
\definecolor {navy}                {rgb}{0.00,0.00,0.50}
\definecolor {cornflowerblue}      {rgb}{0.39,0.58,0.93}
\definecolor {darkslateblue}       {rgb}{0.28,0.24,0.55}
\definecolor {slateblue}           {rgb}{0.42,0.35,0.80}
\definecolor {mediumslateblue}     {rgb}{0.48,0.41,0.93}
\definecolor {lightslateblue}      {rgb}{0.52,0.44,1.00}
\definecolor {mediumblue}          {rgb}{0.00,0.00,0.80}
\definecolor {royalblue}           {rgb}{0.25,0.41,0.88}
\definecolor {blue}                {rgb}{0.00,0.00,1.00}
\definecolor {dodgerblue}          {rgb}{0.12,0.56,1.00}
\definecolor {deepskyblue}         {rgb}{0.00,0.75,1.00}
\definecolor {skyblue}             {rgb}{0.53,0.81,0.92}
\definecolor {lightskyblue}        {rgb}{0.53,0.81,0.98}
\definecolor {steelblue}           {rgb}{0.27,0.51,0.71}
\definecolor {lightsteelblue}      {rgb}{0.69,0.77,0.87}
\definecolor {lightblue}           {rgb}{0.68,0.85,0.90}
\definecolor {powderblue}          {rgb}{0.69,0.88,0.90}
\definecolor {paleturquoise}       {rgb}{0.69,0.93,0.93}
\definecolor {darkturquoise}       {rgb}{0.00,0.81,0.82}
\definecolor {mediumturquoise}     {rgb}{0.28,0.82,0.80}
\definecolor {turquoise}           {rgb}{0.25,0.88,0.82}
\definecolor {cyan}                {rgb}{0.00,1.00,1.00}
\definecolor {lightcyan}           {rgb}{0.88,1.00,1.00}
\definecolor {cadetblue}           {rgb}{0.37,0.62,0.63}
\definecolor {mediumaquamarine}    {rgb}{0.40,0.80,0.67}
\definecolor {aquamarine}          {rgb}{0.50,1.00,0.83}
\definecolor {darkgreen}           {rgb}{0.00,0.39,0.00}
\definecolor {darkolivegreen}      {rgb}{0.33,0.42,0.18}
\definecolor {darkseagreen}        {rgb}{0.56,0.74,0.56}
\definecolor {seagreen}            {rgb}{0.18,0.55,0.34}
\definecolor {mediumseagreen}      {rgb}{0.24,0.70,0.44}
\definecolor {lightseagreen}       {rgb}{0.13,0.70,0.67}
\definecolor {palegreen}           {rgb}{0.60,0.98,0.60}
\definecolor {springgreen}         {rgb}{0.00,1.00,0.50}
\definecolor {lawngreen}           {rgb}{0.49,0.99,0.00}
\definecolor {green}               {rgb}{0.00,1.00,0.00}
\definecolor {chartreuse}          {rgb}{0.50,1.00,0.00}
\definecolor {mediumspringgreen}   {rgb}{0.00,0.98,0.60}
\definecolor {greenyellow}         {rgb}{0.68,1.00,0.18}
\definecolor {limegreen}           {rgb}{0.20,0.80,0.20}
\definecolor {yellowgreen}         {rgb}{0.60,0.80,0.20}
\definecolor {forestgreen}         {rgb}{0.13,0.55,0.13}
\definecolor {olivedrab}           {rgb}{0.42,0.56,0.14}
\definecolor {darkkhaki}           {rgb}{0.74,0.72,0.42}
\definecolor {khaki}               {rgb}{0.94,0.90,0.55}
\definecolor {palegoldenrod}       {rgb}{0.93,0.91,0.67}
\definecolor {lightgoldenrodyellow} {rgb}{0.98,0.98,0.82}
\definecolor {lightyellow}         {rgb}{1.00,1.00,0.88}
\definecolor {yellow}              {rgb}{1.00,1.00,0.00}
\definecolor {gold}                {rgb}{1.00,0.84,0.00}
\definecolor {lightgoldenrod}      {rgb}{0.93,0.87,0.51}
\definecolor {goldenrod}           {rgb}{0.85,0.65,0.13}
\definecolor {darkgoldenrod}       {rgb}{0.72,0.53,0.04}
\definecolor {rosybrown}           {rgb}{0.74,0.56,0.56}
\definecolor {indianred}           {rgb}{0.80,0.36,0.36}
\definecolor {saddlebrown}         {rgb}{0.55,0.27,0.07}
\definecolor {sienna}              {rgb}{0.63,0.32,0.18}
\definecolor {peru}                {rgb}{0.80,0.52,0.25}
\definecolor {burlywood}           {rgb}{0.87,0.72,0.53}
\definecolor {beige}               {rgb}{0.96,0.96,0.86}
\definecolor {wheat}               {rgb}{0.96,0.87,0.70}
\definecolor {sandybrown}          {rgb}{0.96,0.64,0.38}
\definecolor {tan}                 {rgb}{0.82,0.71,0.55}
\definecolor {chocolate}           {rgb}{0.82,0.41,0.12}
\definecolor {firebrick}           {rgb}{0.70,0.13,0.13}
\definecolor {brown}               {rgb}{0.65,0.16,0.16}
\definecolor {darksalmon}          {rgb}{0.91,0.59,0.48}
\definecolor {salmon}              {rgb}{0.98,0.50,0.45}
\definecolor {lightsalmon}         {rgb}{1.00,0.63,0.48}
\definecolor {orange}              {rgb}{1.00,0.65,0.00}
\definecolor {darkorange}          {rgb}{1.00,0.55,0.00}
\definecolor {coral}               {rgb}{1.00,0.50,0.31}
\definecolor {lightcoral}          {rgb}{0.94,0.50,0.50}
\definecolor {tomato}              {rgb}{1.00,0.39,0.28}
\definecolor {orangered}           {rgb}{1.00,0.27,0.00}
\definecolor {red}                 {rgb}{1.00,0.00,0.00}
\definecolor {hotpink}             {rgb}{1.00,0.41,0.71}
\definecolor {deeppink}            {rgb}{1.00,0.08,0.58}
\definecolor {pink}                {rgb}{1.00,0.75,0.80}
\definecolor {lightpink}           {rgb}{1.00,0.71,0.76}
\definecolor {palevioletred}       {rgb}{0.86,0.44,0.58}
\definecolor {maroon}              {rgb}{0.69,0.19,0.38}
\definecolor {mediumvioletred}     {rgb}{0.78,0.08,0.52}
\definecolor {violetred}           {rgb}{0.82,0.13,0.56}
\definecolor {magenta}             {rgb}{1.00,0.00,1.00}
\definecolor {violet}              {rgb}{0.93,0.51,0.93}
\definecolor {plum}                {rgb}{0.87,0.63,0.87}
\definecolor {orchid}              {rgb}{0.85,0.44,0.84}
\definecolor {mediumorchid}        {rgb}{0.73,0.33,0.83}
\definecolor {darkorchid}          {rgb}{0.60,0.20,0.80}
\definecolor {darkviolet}          {rgb}{0.58,0.00,0.83}
\definecolor {blueviolet}          {rgb}{0.54,0.17,0.89}
\definecolor {purple}              {rgb}{0.63,0.13,0.94}
\definecolor {mediumpurple}        {rgb}{0.58,0.44,0.86}
\definecolor {thistle}             {rgb}{0.85,0.75,0.85}
\definecolor {snow2}               {rgb}{0.93,0.91,0.91}
\definecolor {snow3}               {rgb}{0.80,0.79,0.79}
\definecolor {snow4}               {rgb}{0.55,0.54,0.54}
\definecolor {seashell2}           {rgb}{0.93,0.90,0.87}
\definecolor {seashell3}           {rgb}{0.80,0.77,0.75}
\definecolor {seashell4}           {rgb}{0.55,0.53,0.51}
\definecolor {antiquewhite1}       {rgb}{1.00,0.94,0.86}
\definecolor {antiquewhite2}       {rgb}{0.93,0.87,0.80}
\definecolor {antiquewhite3}       {rgb}{0.80,0.75,0.69}
\definecolor {antiquewhite4}       {rgb}{0.55,0.51,0.47}
\definecolor {bisque2}             {rgb}{0.93,0.84,0.72}
\definecolor {bisque3}             {rgb}{0.80,0.72,0.62}
\definecolor {bisque4}             {rgb}{0.55,0.49,0.42}
\definecolor {peachpuff2}          {rgb}{0.93,0.80,0.68}
\definecolor {peachpuff3}          {rgb}{0.80,0.69,0.58}
\definecolor {peachpuff4}          {rgb}{0.55,0.47,0.40}
\definecolor {navajowhite2}        {rgb}{0.93,0.81,0.63}
\definecolor {navajowhite3}        {rgb}{0.80,0.70,0.55}
\definecolor {navajowhite4}        {rgb}{0.55,0.47,0.37}
\definecolor {lemonchiffon2}       {rgb}{0.93,0.91,0.75}
\definecolor {lemonchiffon3}       {rgb}{0.80,0.79,0.65}
\definecolor {lemonchiffon4}       {rgb}{0.55,0.54,0.44}
\definecolor {cornsilk2}           {rgb}{0.93,0.91,0.80}
\definecolor {cornsilk3}           {rgb}{0.80,0.78,0.69}
\definecolor {cornsilk4}           {rgb}{0.55,0.53,0.47}
\definecolor {ivory2}              {rgb}{0.93,0.93,0.88}
\definecolor {ivory3}              {rgb}{0.80,0.80,0.76}
\definecolor {ivory4}              {rgb}{0.55,0.55,0.51}
\definecolor {honeydew2}           {rgb}{0.88,0.93,0.88}
\definecolor {honeydew3}           {rgb}{0.76,0.80,0.76}
\definecolor {honeydew4}           {rgb}{0.51,0.55,0.51}
\definecolor {lavenderblush2}      {rgb}{0.93,0.88,0.90}
\definecolor {lavenderblush3}      {rgb}{0.80,0.76,0.77}
\definecolor {lavenderblush4}      {rgb}{0.55,0.51,0.53}
\definecolor {mistyrose2}          {rgb}{0.93,0.84,0.82}
\definecolor {mistyrose3}          {rgb}{0.80,0.72,0.71}
\definecolor {mistyrose4}          {rgb}{0.55,0.49,0.48}
\definecolor {azure2}              {rgb}{0.88,0.93,0.93}
\definecolor {azure3}              {rgb}{0.76,0.80,0.80}
\definecolor {azure4}              {rgb}{0.51,0.55,0.55}
\definecolor {slateblue1}          {rgb}{0.51,0.44,1.00}
\definecolor {slateblue2}          {rgb}{0.48,0.40,0.93}
\definecolor {slateblue3}          {rgb}{0.41,0.35,0.80}
\definecolor {slateblue4}          {rgb}{0.28,0.24,0.55}
\definecolor {royalblue1}          {rgb}{0.28,0.46,1.00}
\definecolor {royalblue2}          {rgb}{0.26,0.43,0.93}
\definecolor {royalblue3}          {rgb}{0.23,0.37,0.80}
\definecolor {royalblue4}          {rgb}{0.15,0.25,0.55}
\definecolor {blue2}               {rgb}{0.00,0.00,0.93}
\definecolor {blue4}               {rgb}{0.00,0.00,0.55}
\definecolor {dodgerblue2}         {rgb}{0.11,0.53,0.93}
\definecolor {dodgerblue3}         {rgb}{0.09,0.45,0.80}
\definecolor {dodgerblue4}         {rgb}{0.06,0.31,0.55}
\definecolor {steelblue1}          {rgb}{0.39,0.72,1.00}
\definecolor {steelblue2}          {rgb}{0.36,0.67,0.93}
\definecolor {steelblue3}          {rgb}{0.31,0.58,0.80}
\definecolor {steelblue4}          {rgb}{0.21,0.39,0.55}
\definecolor {deepskyblue2}        {rgb}{0.00,0.70,0.93}
\definecolor {deepskyblue3}        {rgb}{0.00,0.60,0.80}
\definecolor {deepskyblue4}        {rgb}{0.00,0.41,0.55}
\definecolor {skyblue1}            {rgb}{0.53,0.81,1.00}
\definecolor {skyblue2}            {rgb}{0.49,0.75,0.93}
\definecolor {skyblue3}            {rgb}{0.42,0.65,0.80}
\definecolor {skyblue4}            {rgb}{0.29,0.44,0.55}
\definecolor {lightskyblue1}       {rgb}{0.69,0.89,1.00}
\definecolor {lightskyblue2}       {rgb}{0.64,0.83,0.93}
\definecolor {lightskyblue3}       {rgb}{0.55,0.71,0.80}
\definecolor {lightskyblue4}       {rgb}{0.38,0.48,0.55}
\definecolor {slategray1}          {rgb}{0.78,0.89,1.00}
\definecolor {slategray2}          {rgb}{0.73,0.83,0.93}
\definecolor {slategray3}          {rgb}{0.62,0.71,0.80}
\definecolor {slategray4}          {rgb}{0.42,0.48,0.55}
\definecolor {lightsteelblue1}     {rgb}{0.79,0.88,1.00}
\definecolor {lightsteelblue2}     {rgb}{0.74,0.82,0.93}
\definecolor {lightsteelblue3}     {rgb}{0.64,0.71,0.80}
\definecolor {lightsteelblue4}     {rgb}{0.43,0.48,0.55}
\definecolor {lightblue1}          {rgb}{0.75,0.94,1.00}
\definecolor {lightblue2}          {rgb}{0.70,0.87,0.93}
\definecolor {lightblue3}          {rgb}{0.60,0.75,0.80}
\definecolor {lightblue4}          {rgb}{0.41,0.51,0.55}
\definecolor {lightcyan2}          {rgb}{0.82,0.93,0.93}
\definecolor {lightcyan3}          {rgb}{0.71,0.80,0.80}
\definecolor {lightcyan4}          {rgb}{0.48,0.55,0.55}
\definecolor {paleturquoise1}      {rgb}{0.73,1.00,1.00}
\definecolor {paleturquoise2}      {rgb}{0.68,0.93,0.93}
\definecolor {paleturquoise3}      {rgb}{0.59,0.80,0.80}
\definecolor {paleturquoise4}      {rgb}{0.40,0.55,0.55}
\definecolor {cadetblue1}          {rgb}{0.60,0.96,1.00}
\definecolor {cadetblue2}          {rgb}{0.56,0.90,0.93}
\definecolor {cadetblue3}          {rgb}{0.48,0.77,0.80}
\definecolor {cadetblue4}          {rgb}{0.33,0.53,0.55}
\definecolor {turquoise1}          {rgb}{0.00,0.96,1.00}
\definecolor {turquoise2}          {rgb}{0.00,0.90,0.93}
\definecolor {turquoise3}          {rgb}{0.00,0.77,0.80}
\definecolor {turquoise4}          {rgb}{0.00,0.53,0.55}
\definecolor {cyan2}               {rgb}{0.00,0.93,0.93}
\definecolor {cyan3}               {rgb}{0.00,0.80,0.80}
\definecolor {cyan4}               {rgb}{0.00,0.55,0.55}
\definecolor {darkslategray1}      {rgb}{0.59,1.00,1.00}
\definecolor {darkslategray2}      {rgb}{0.55,0.93,0.93}
\definecolor {darkslategray3}      {rgb}{0.47,0.80,0.80}
\definecolor {darkslategray4}      {rgb}{0.32,0.55,0.55}
\definecolor {aquamarine2}         {rgb}{0.46,0.93,0.78}
\definecolor {aquamarine4}         {rgb}{0.27,0.55,0.45}
\definecolor {darkseagreen1}       {rgb}{0.76,1.00,0.76}
\definecolor {darkseagreen2}       {rgb}{0.71,0.93,0.71}
\definecolor {darkseagreen3}       {rgb}{0.61,0.80,0.61}
\definecolor {darkseagreen4}       {rgb}{0.41,0.55,0.41}
\definecolor {seagreen1}           {rgb}{0.33,1.00,0.62}
\definecolor {seagreen2}           {rgb}{0.31,0.93,0.58}
\definecolor {seagreen3}           {rgb}{0.26,0.80,0.50}
\definecolor {palegreen1}          {rgb}{0.60,1.00,0.60}
\definecolor {palegreen2}          {rgb}{0.56,0.93,0.56}
\definecolor {palegreen3}          {rgb}{0.49,0.80,0.49}
\definecolor {palegreen4}          {rgb}{0.33,0.55,0.33}
\definecolor {springgreen2}        {rgb}{0.00,0.93,0.46}
\definecolor {springgreen3}        {rgb}{0.00,0.80,0.40}
\definecolor {springgreen4}        {rgb}{0.00,0.55,0.27}
\definecolor {green2}              {rgb}{0.00,0.93,0.00}
\definecolor {green3}              {rgb}{0.00,0.80,0.00}
\definecolor {green4}              {rgb}{0.00,0.55,0.00}
\definecolor {chartreuse2}         {rgb}{0.46,0.93,0.00}
\definecolor {chartreuse3}         {rgb}{0.40,0.80,0.00}
\definecolor {chartreuse4}         {rgb}{0.27,0.55,0.00}
\definecolor {olivedrab1}          {rgb}{0.75,1.00,0.24}
\definecolor {olivedrab2}          {rgb}{0.70,0.93,0.23}
\definecolor {olivedrab4}          {rgb}{0.41,0.55,0.13}
\definecolor {darkolivegreen1}     {rgb}{0.79,1.00,0.44}
\definecolor {darkolivegreen2}     {rgb}{0.74,0.93,0.41}
\definecolor {darkolivegreen3}     {rgb}{0.64,0.80,0.35}
\definecolor {darkolivegreen4}     {rgb}{0.43,0.55,0.24}
\definecolor {khaki1}              {rgb}{1.00,0.96,0.56}
\definecolor {khaki2}              {rgb}{0.93,0.90,0.52}
\definecolor {khaki3}              {rgb}{0.80,0.78,0.45}
\definecolor {khaki4}              {rgb}{0.55,0.53,0.31}
\definecolor {lightgoldenrod1}     {rgb}{1.00,0.93,0.55}
\definecolor {lightgoldenrod2}     {rgb}{0.93,0.86,0.51}
\definecolor {lightgoldenrod3}     {rgb}{0.80,0.75,0.44}
\definecolor {lightgoldenrod4}     {rgb}{0.55,0.51,0.30}
\definecolor {lightyellow2}        {rgb}{0.93,0.93,0.82}
\definecolor {lightyellow3}        {rgb}{0.80,0.80,0.71}
\definecolor {lightyellow4}        {rgb}{0.55,0.55,0.48}
\definecolor {yellow2}             {rgb}{0.93,0.93,0.00}
\definecolor {yellow3}             {rgb}{0.80,0.80,0.00}
\definecolor {yellow4}             {rgb}{0.55,0.55,0.00}
\definecolor {gold2}               {rgb}{0.93,0.79,0.00}
\definecolor {gold3}               {rgb}{0.80,0.68,0.00}
\definecolor {gold4}               {rgb}{0.55,0.46,0.00}
\definecolor {goldenrod1}          {rgb}{1.00,0.76,0.15}
\definecolor {goldenrod2}          {rgb}{0.93,0.71,0.13}
\definecolor {goldenrod3}          {rgb}{0.80,0.61,0.11}
\definecolor {goldenrod4}          {rgb}{0.55,0.41,0.08}
\definecolor {darkgoldenrod1}      {rgb}{1.00,0.73,0.06}
\definecolor {darkgoldenrod2}      {rgb}{0.93,0.68,0.05}
\definecolor {darkgoldenrod3}      {rgb}{0.80,0.58,0.05}
\definecolor {darkgoldenrod4}      {rgb}{0.55,0.40,0.03}
\definecolor {rosybrown1}          {rgb}{1.00,0.76,0.76}
\definecolor {rosybrown2}          {rgb}{0.93,0.71,0.71}
\definecolor {rosybrown3}          {rgb}{0.80,0.61,0.61}
\definecolor {rosybrown4}          {rgb}{0.55,0.41,0.41}
\definecolor {indianred1}          {rgb}{1.00,0.42,0.42}
\definecolor {indianred2}          {rgb}{0.93,0.39,0.39}
\definecolor {indianred3}          {rgb}{0.80,0.33,0.33}
\definecolor {indianred4}          {rgb}{0.55,0.23,0.23}
\definecolor {sienna1}             {rgb}{1.00,0.51,0.28}
\definecolor {sienna2}             {rgb}{0.93,0.47,0.26}
\definecolor {sienna3}             {rgb}{0.80,0.41,0.22}
\definecolor {sienna4}             {rgb}{0.55,0.28,0.15}
\definecolor {burlywood1}          {rgb}{1.00,0.83,0.61}
\definecolor {burlywood2}          {rgb}{0.93,0.77,0.57}
\definecolor {burlywood3}          {rgb}{0.80,0.67,0.49}
\definecolor {burlywood4}          {rgb}{0.55,0.45,0.33}
\definecolor {wheat1}              {rgb}{1.00,0.91,0.73}
\definecolor {wheat2}              {rgb}{0.93,0.85,0.68}
\definecolor {wheat3}              {rgb}{0.80,0.73,0.59}
\definecolor {wheat4}              {rgb}{0.55,0.49,0.40}
\definecolor {tan1}                {rgb}{1.00,0.65,0.31}
\definecolor {tan2}                {rgb}{0.93,0.60,0.29}
\definecolor {tan4}                {rgb}{0.55,0.35,0.17}
\definecolor {chocolate1}          {rgb}{1.00,0.50,0.14}
\definecolor {chocolate2}          {rgb}{0.93,0.46,0.13}
\definecolor {chocolate3}          {rgb}{0.80,0.40,0.11}
\definecolor {firebrick1}          {rgb}{1.00,0.19,0.19}
\definecolor {firebrick2}          {rgb}{0.93,0.17,0.17}
\definecolor {firebrick3}          {rgb}{0.80,0.15,0.15}
\definecolor {firebrick4}          {rgb}{0.55,0.10,0.10}
\definecolor {brown1}              {rgb}{1.00,0.25,0.25}
\definecolor {brown2}              {rgb}{0.93,0.23,0.23}
\definecolor {brown3}              {rgb}{0.80,0.20,0.20}
\definecolor {brown4}              {rgb}{0.55,0.14,0.14}
\definecolor {salmon1}             {rgb}{1.00,0.55,0.41}
\definecolor {salmon2}             {rgb}{0.93,0.51,0.38}
\definecolor {salmon3}             {rgb}{0.80,0.44,0.33}
\definecolor {salmon4}             {rgb}{0.55,0.30,0.22}
\definecolor {lightsalmon2}        {rgb}{0.93,0.58,0.45}
\definecolor {lightsalmon3}        {rgb}{0.80,0.51,0.38}
\definecolor {lightsalmon4}        {rgb}{0.55,0.34,0.26}
\definecolor {orange2}             {rgb}{0.93,0.60,0.00}
\definecolor {orange3}             {rgb}{0.80,0.52,0.00}
\definecolor {orange4}             {rgb}{0.55,0.35,0.00}
\definecolor {darkorange1}         {rgb}{1.00,0.50,0.00}
\definecolor {darkorange2}         {rgb}{0.93,0.46,0.00}
\definecolor {darkorange3}         {rgb}{0.80,0.40,0.00}
\definecolor {darkorange4}         {rgb}{0.55,0.27,0.00}
\definecolor {coral1}              {rgb}{1.00,0.45,0.34}
\definecolor {coral2}              {rgb}{0.93,0.42,0.31}
\definecolor {coral3}              {rgb}{0.80,0.36,0.27}
\definecolor {coral4}              {rgb}{0.55,0.24,0.18}
\definecolor {tomato2}             {rgb}{0.93,0.36,0.26}
\definecolor {tomato3}             {rgb}{0.80,0.31,0.22}
\definecolor {tomato4}             {rgb}{0.55,0.21,0.15}
\definecolor {orangered2}          {rgb}{0.93,0.25,0.00}
\definecolor {orangered3}          {rgb}{0.80,0.22,0.00}
\definecolor {orangered4}          {rgb}{0.55,0.15,0.00}
\definecolor {red2}                {rgb}{0.93,0.00,0.00}
\definecolor {red3}                {rgb}{0.80,0.00,0.00}
\definecolor {red4}                {rgb}{0.55,0.00,0.00}
\definecolor {deeppink2}           {rgb}{0.93,0.07,0.54}
\definecolor {deeppink3}           {rgb}{0.80,0.06,0.46}
\definecolor {deeppink4}           {rgb}{0.55,0.04,0.31}
\definecolor {hotpink1}            {rgb}{1.00,0.43,0.71}
\definecolor {hotpink2}            {rgb}{0.93,0.42,0.65}
\definecolor {hotpink3}            {rgb}{0.80,0.38,0.56}
\definecolor {hotpink4}            {rgb}{0.55,0.23,0.38}
\definecolor {pink1}               {rgb}{1.00,0.71,0.77}
\definecolor {pink2}               {rgb}{0.93,0.66,0.72}
\definecolor {pink3}               {rgb}{0.80,0.57,0.62}
\definecolor {pink4}               {rgb}{0.55,0.39,0.42}
\definecolor {lightpink1}          {rgb}{1.00,0.68,0.73}
\definecolor {lightpink2}          {rgb}{0.93,0.64,0.68}
\definecolor {lightpink3}          {rgb}{0.80,0.55,0.58}
\definecolor {lightpink4}          {rgb}{0.55,0.37,0.40}
\definecolor {palevioletred1}      {rgb}{1.00,0.51,0.67}
\definecolor {palevioletred2}      {rgb}{0.93,0.47,0.62}
\definecolor {palevioletred3}      {rgb}{0.80,0.41,0.54}
\definecolor {palevioletred4}      {rgb}{0.55,0.28,0.36}
\definecolor {maroon1}             {rgb}{1.00,0.20,0.70}
\definecolor {maroon2}             {rgb}{0.93,0.19,0.65}
\definecolor {maroon3}             {rgb}{0.80,0.16,0.56}
\definecolor {maroon4}             {rgb}{0.55,0.11,0.38}
\definecolor {violetred1}          {rgb}{1.00,0.24,0.59}
\definecolor {violetred2}          {rgb}{0.93,0.23,0.55}
\definecolor {violetred3}          {rgb}{0.80,0.20,0.47}
\definecolor {violetred4}          {rgb}{0.55,0.13,0.32}
\definecolor {magenta2}            {rgb}{0.93,0.00,0.93}
\definecolor {magenta3}            {rgb}{0.80,0.00,0.80}
\definecolor {magenta4}            {rgb}{0.55,0.00,0.55}
\definecolor {orchid1}             {rgb}{1.00,0.51,0.98}
\definecolor {orchid2}             {rgb}{0.93,0.48,0.91}
\definecolor {orchid3}             {rgb}{0.80,0.41,0.79}
\definecolor {orchid4}             {rgb}{0.55,0.28,0.54}
\definecolor {plum1}               {rgb}{1.00,0.73,1.00}
\definecolor {plum2}               {rgb}{0.93,0.68,0.93}
\definecolor {plum3}               {rgb}{0.80,0.59,0.80}
\definecolor {plum4}               {rgb}{0.55,0.40,0.55}
\definecolor {mediumorchid1}       {rgb}{0.88,0.40,1.00}
\definecolor {mediumorchid2}       {rgb}{0.82,0.37,0.93}
\definecolor {mediumorchid3}       {rgb}{0.71,0.32,0.80}
\definecolor {mediumorchid4}       {rgb}{0.48,0.22,0.55}
\definecolor {darkorchid1}         {rgb}{0.75,0.24,1.00}
\definecolor {darkorchid2}         {rgb}{0.70,0.23,0.93}
\definecolor {darkorchid3}         {rgb}{0.60,0.20,0.80}
\definecolor {darkorchid4}         {rgb}{0.41,0.13,0.55}
\definecolor {purple1}             {rgb}{0.61,0.19,1.00}
\definecolor {purple2}             {rgb}{0.57,0.17,0.93}
\definecolor {purple3}             {rgb}{0.49,0.15,0.80}
\definecolor {purple4}             {rgb}{0.33,0.10,0.55}
\definecolor {mediumpurple1}       {rgb}{0.67,0.51,1.00}
\definecolor {mediumpurple2}       {rgb}{0.62,0.47,0.93}
\definecolor {mediumpurple3}       {rgb}{0.54,0.41,0.80}
\definecolor {mediumpurple4}       {rgb}{0.36,0.28,0.55}
\definecolor {thistle1}            {rgb}{1.00,0.88,1.00}
\definecolor {thistle2}            {rgb}{0.93,0.82,0.93}
\definecolor {thistle3}            {rgb}{0.80,0.71,0.80}
\definecolor {thistle4}            {rgb}{0.55,0.48,0.55}
\definecolor {gray1}               {rgb}{0.01,0.01,0.01}
\definecolor {gray2}               {rgb}{0.02,0.02,0.02}
\definecolor {gray3}               {rgb}{0.03,0.03,0.03}
\definecolor {gray4}               {rgb}{0.04,0.04,0.04}
\definecolor {gray5}               {rgb}{0.05,0.05,0.05}
\definecolor {gray6}               {rgb}{0.06,0.06,0.06}
\definecolor {gray7}               {rgb}{0.07,0.07,0.07}
\definecolor {gray8}               {rgb}{0.08,0.08,0.08}
\definecolor {gray9}               {rgb}{0.09,0.09,0.09}
\definecolor {gray10}              {rgb}{0.10,0.10,0.10}
\definecolor {gray11}              {rgb}{0.11,0.11,0.11}
\definecolor {gray12}              {rgb}{0.12,0.12,0.12}
\definecolor {gray13}              {rgb}{0.13,0.13,0.13}
\definecolor {gray14}              {rgb}{0.14,0.14,0.14}
\definecolor {gray15}              {rgb}{0.15,0.15,0.15}
\definecolor {gray16}              {rgb}{0.16,0.16,0.16}
\definecolor {gray17}              {rgb}{0.17,0.17,0.17}
\definecolor {gray18}              {rgb}{0.18,0.18,0.18}
\definecolor {gray19}              {rgb}{0.19,0.19,0.19}
\definecolor {gray20}              {rgb}{0.20,0.20,0.20}
\definecolor {gray21}              {rgb}{0.21,0.21,0.21}
\definecolor {gray22}              {rgb}{0.22,0.22,0.22}
\definecolor {gray23}              {rgb}{0.23,0.23,0.23}
\definecolor {gray24}              {rgb}{0.24,0.24,0.24}
\definecolor {gray25}              {rgb}{0.25,0.25,0.25}
\definecolor {gray26}              {rgb}{0.26,0.26,0.26}
\definecolor {gray27}              {rgb}{0.27,0.27,0.27}
\definecolor {gray28}              {rgb}{0.28,0.28,0.28}
\definecolor {gray29}              {rgb}{0.29,0.29,0.29}
\definecolor {gray30}              {rgb}{0.30,0.30,0.30}
\definecolor {gray31}              {rgb}{0.31,0.31,0.31}
\definecolor {gray32}              {rgb}{0.32,0.32,0.32}
\definecolor {gray33}              {rgb}{0.33,0.33,0.33}
\definecolor {gray34}              {rgb}{0.34,0.34,0.34}
\definecolor {gray35}              {rgb}{0.35,0.35,0.35}
\definecolor {gray36}              {rgb}{0.36,0.36,0.36}
\definecolor {gray37}              {rgb}{0.37,0.37,0.37}
\definecolor {gray38}              {rgb}{0.38,0.38,0.38}
\definecolor {gray39}              {rgb}{0.39,0.39,0.39}
\definecolor {gray40}              {rgb}{0.40,0.40,0.40}
\definecolor {gray42}              {rgb}{0.42,0.42,0.42}
\definecolor {gray43}              {rgb}{0.43,0.43,0.43}
\definecolor {gray44}              {rgb}{0.44,0.44,0.44}
\definecolor {gray45}              {rgb}{0.45,0.45,0.45}
\definecolor {gray46}              {rgb}{0.46,0.46,0.46}
\definecolor {gray47}              {rgb}{0.47,0.47,0.47}
\definecolor {gray48}              {rgb}{0.48,0.48,0.48}
\definecolor {gray49}              {rgb}{0.49,0.49,0.49}
\definecolor {gray50}              {rgb}{0.50,0.50,0.50}
\definecolor {gray51}              {rgb}{0.51,0.51,0.51}
\definecolor {gray52}              {rgb}{0.52,0.52,0.52}
\definecolor {gray53}              {rgb}{0.53,0.53,0.53}
\definecolor {gray54}              {rgb}{0.54,0.54,0.54}
\definecolor {gray55}              {rgb}{0.55,0.55,0.55}
\definecolor {gray56}              {rgb}{0.56,0.56,0.56}
\definecolor {gray57}              {rgb}{0.57,0.57,0.57}
\definecolor {gray58}              {rgb}{0.58,0.58,0.58}
\definecolor {gray59}              {rgb}{0.59,0.59,0.59}
\definecolor {gray60}              {rgb}{0.60,0.60,0.60}
\definecolor {gray61}              {rgb}{0.61,0.61,0.61}
\definecolor {gray62}              {rgb}{0.62,0.62,0.62}
\definecolor {gray63}              {rgb}{0.63,0.63,0.63}
\definecolor {gray64}              {rgb}{0.64,0.64,0.64}
\definecolor {gray65}              {rgb}{0.65,0.65,0.65}
\definecolor {gray66}              {rgb}{0.66,0.66,0.66}
\definecolor {gray67}              {rgb}{0.67,0.67,0.67}
\definecolor {gray68}              {rgb}{0.68,0.68,0.68}
\definecolor {gray69}              {rgb}{0.69,0.69,0.69}
\definecolor {gray70}              {rgb}{0.70,0.70,0.70}
\definecolor {gray71}              {rgb}{0.71,0.71,0.71}
\definecolor {gray72}              {rgb}{0.72,0.72,0.72}
\definecolor {gray73}              {rgb}{0.73,0.73,0.73}
\definecolor {gray74}              {rgb}{0.74,0.74,0.74}
\definecolor {gray75}              {rgb}{0.75,0.75,0.75}
\definecolor {gray76}              {rgb}{0.76,0.76,0.76}
\definecolor {gray77}              {rgb}{0.77,0.77,0.77}
\definecolor {gray78}              {rgb}{0.78,0.78,0.78}
\definecolor {gray79}              {rgb}{0.79,0.79,0.79}
\definecolor {gray80}              {rgb}{0.80,0.80,0.80}
\definecolor {gray81}              {rgb}{0.81,0.81,0.81}
\definecolor {gray82}              {rgb}{0.82,0.82,0.82}
\definecolor {gray83}              {rgb}{0.83,0.83,0.83}
\definecolor {gray84}              {rgb}{0.84,0.84,0.84}
\definecolor {gray85}              {rgb}{0.85,0.85,0.85}
\definecolor {gray86}              {rgb}{0.86,0.86,0.86}
\definecolor {gray87}              {rgb}{0.87,0.87,0.87}
\definecolor {gray88}              {rgb}{0.88,0.88,0.88}
\definecolor {gray89}              {rgb}{0.89,0.89,0.89}
\definecolor {gray90}              {rgb}{0.90,0.90,0.90}
\definecolor {gray91}              {rgb}{0.91,0.91,0.91}
\definecolor {gray92}              {rgb}{0.92,0.92,0.92}
\definecolor {gray93}              {rgb}{0.93,0.93,0.93}
\definecolor {gray94}              {rgb}{0.94,0.94,0.94}
\definecolor {gray95}              {rgb}{0.95,0.95,0.95}
\definecolor {gray97}              {rgb}{0.97,0.97,0.97}
\definecolor {gray98}              {rgb}{0.98,0.98,0.98}
\definecolor {gray99}              {rgb}{0.99,0.99,0.99}
\definecolor {darkgrey}            {rgb}{0.66,0.66,0.66}

\newcommand{\ignore}[1]{}

\newenvironment{gschange}{\color{black}}{\normalcolor}

\newcommand{\ignoreinshort}[1]{}
 
\def\makenewenumerate#1#2{%
\newcounter{cnt#1}
\newenvironment{#1}%
{\begin{list}{\makebox[0pt][r]{#2}}%
{\setlength{\itemsep}{0pt}%
 \setlength{\parsep}{.2em}%
 \setlength{\leftmargin}{1.5em}%
 \setlength{\labelwidth}{.4em}%
 \usecounter{cnt#1}}}
{\end{list}}}

\makenewenumerate{myenumerate}{\arabic{cntmyenumerate}.}

\makenewenumerate{renumerate}{\rm(\roman{cntrenumerate})}
\makenewenumerate{renumerateprime}{\rm(\roman{cntrenumerateprime}$'$)}
\makenewenumerate{renumeratesecond}{\rm(\roman{cntrenumeratesecond}$''$)}

\makenewenumerate{aenumerate}{({\it\alph{cntaenumerate}})}
\makenewenumerate{aenumerateprime}{({\it\alph{cntaenumerateprime}$'$})}
\makenewenumerate{aenumeratesecond}{({\it\alph{cntaenumeratesecond}$''$})}

\def\newplaintheorem#1#2{%
\newtheorem{#1plain}{#2}
\newenvironment{#1}{\begin{#1plain}\rm }{\end{#1plain}}}

\newplaintheorem{notation}{Notation}
\newplaintheorem{cexample}{Counter-Example}

\newcommand{\sref}[1]{\S{}\ref{#1}}

\newcommand{\pos}{\phantom{\neg}}

\newcommand{\solver}[1]{{\sc TabularAllSAT$_{AAAI24}${#1}}\xspace}
\newcommand{\solverPlus}[1]{{\sc TabularAllSAT{#1}}\xspace}
\newcommand{\solverSMT}[1]{{\sc TabularAllSMT{#1}}\xspace}

\newcommand{\vicnf}{\ensuremath{F_{CNF}}}

\newcommand\mysout{\bgroup \markoverwith{{-}}\ULon}
\newcommand\nosout{\bgroup \markoverwith{{ }}\ULon}
\definecolor{mygray}{rgb}{0.90,0.90,0.90}
\definecolor{mywhite}{rgb}{1.00,1.00,1.00}

\newcommand{\T}{\ensuremath{\mathcal{T}}\xspace}

\newcommand{\smtt}{\ensuremath{\text{SMT}(\T)}\xspace}

\newcommand{\allA}{\ensuremath{\mathbf{A}}\xspace}
\newcommand{\allB}{\ensuremath{\mathbf{B}}\xspace}

\newcommand{\ti}[1]{\ensuremath{\sf{t}^{(#1)}}\xspace}
\newcommand{\tn}[1]{\ti{n}}

\newcommand{\GMCHANGE}[1]{\textcolor{black}{#1}}

\newcommand{\GSCHANGEBIS}[1]{\textcolor{black}{#1}}

\newcommand{\blue}[1]{\textcolor{blue}{#1}}

\newcommand{\red}[1]{\textcolor{red}{#1}}
\newcommand{\green}[1]{\textcolor{darkgreen}{#1}}

\newcommand{\OptFN}[2]{{\ifx&#2&\ensuremath{#1}\else\ensuremath{#1(#2)}\fi}}

\newcommand{\exdone}{\ensuremath{\hfill\diamond}}
\newcommand{\trail}{\emph{T}}
\newcommand{\simplify}{{\sc Check-Literal}}

\newtheorem{definition}{Definition}
\newtheorem{example}{Example}

\begin{document}


\begin{frontmatter}



\title{Disjoint Projected Enumeration for SAT and SMT\\without Blocking Clauses}


\author[label1]{Giuseppe Spallitta} 
\affiliation[label1]{organization={University of Trento},
            addressline={Via Sommarive 9},
            city={Trento},
            postcode={38123},
            state={Italy},
            country={}}

\author[label1]{Roberto Sebastiani} 

\author[label2]{Armin Biere} 
\affiliation[label2]{organization={University of Freiburg},
            addressline={Fahnenbergplatz},
            city={Freiburg},
            postcode={79085},
            state={Germany},
            country={}}

\begin{abstract}
All-Solution Satisfiability (AllSAT) and its extension, All-Solution Satisfiability Modulo Theories (AllSMT), have become more relevant in recent years, mainly in formal verification and artificial intelligence applications. The goal of these problems is the enumeration of all satisfying assignments of a formula (for SAT and SMT problems, respectively), making them useful for test generation, model checking, and probabilistic inference. Nevertheless, traditional AllSAT algorithms face significant computational challenges due to the exponential growth of the search space and inefficiencies caused by blocking clauses, which cause memory blowups and degrade unit propagation performances in the long term. This paper presents two novel solvers: \solverPlus{}, a projected AllSAT solver, and \solverSMT{}, a projected AllSMT solver. Both solvers combine Conflict-Driven Clause Learning (CDCL) with chronological backtracking to improve efficiency while ensuring disjoint enumeration. To retrieve compact partial assignments we propose a novel aggressive implicant shrinking algorithm, compatible with chronological backtracking, to minimize the number of partial assignments, reducing overall search complexity. Furthermore, we extend the solver framework to handle projected enumeration and SMT formulas effectively and efficiently, adapting the baseline framework to integrate theory reasoning and the distinction between important and non-important variables. An extensive experimental evaluation demonstrates the superiority of our approach compared to state-of-the-art solvers, particularly in scenarios requiring projection and SMT-based reasoning.
 \end{abstract}




\end{frontmatter}



\section{Introduction}
\label{sec:introduction}

\begin{gschange}
Given a propositional formula $F$ over a set of Boolean variables $V$, the All-Solution Satisfiability (AllSAT) problem consists of identifying all possible satisfying assignments for $F$. AllSAT has seen significant use across various fields, particularly in hardware and software verification. For example, it has been applied in the automatic generation of program test suites \cite{khurshid2004case}, as well as in both bounded and unbounded model checking \cite{jin2005efficient}. Additionally, AllSAT has been employed in data mining, specifically in solving the frequent itemset mining problem \cite{dlala2016comparative}.

AllSAT can be extended to richer logical frameworks in the form of AllSMT (All Satisfiability Modulo Theories). Whereas AllSAT focuses on Boolean variables and propositional formulas, AllSMT expands to formulas $F$ with variables from more complex domains, interpreted over one or multiple specific first-order logic theories $\mathcal{T}$ (e.g., linear integer arithmetic ($\mathcal{L}\mathcal{I}\mathcal{A}$), linear real arithmetic ($\mathcal{L}\mathcal{R}\mathcal{A}$) or bit-vectors ($\mathcal{B}\mathcal{V}$)). The goal in AllSMT is to enumerate all satisfying assignments $\eta$ for $F$ under the constraints of the theory $\mathcal{T}$. Recently, AllSMT has gained interest in artificial intelligence and has been used for tasks such as probabilistic inference in hybrid domains via Weighted Model Integration (WMI) \cite{spallitta2022smt, spallitta2024enhancing}. Moreover, model counting over first-order theories (\#SMT) \cite{chistikov2015approximate} and the extraction of theory lemmas to generate canonical decision diagrams modulo theories \cite{michelutti2024canonical} rely on enumeration strategies from AllSMT.

In both AllSAT and AllSMT, the concept of \textit{projection} plays an essential role. It involves restricting the enumeration to a specific subset $V_r$ of the variables in $V$, thereby simplifying the search of satisfying models by ignoring the truth value of variables outside of $V_r$.  Projection is particularly beneficial for the enumeration of non-CNF formulas. When formulas are transformed into CNF---often via the Tseitin transformation---projection is essential to exclude the newly introduced auxiliary variables. Projection has been applied in numerous domains other than enumeration, including predicate abstraction \cite{lahiri2006smt}, image computation \cite{gupta2000sat, grumberg2004memory}, quantifier elimination \cite{brauer2011existential}, and model checking \cite{shtrichman2001pruning}.

\end{gschange}

\paragraph{Computational Challenges in Enumeration}
Enumerating all solutions of a given formula $F$ is a significantly more computationally intensive task than solving a single SAT instance. When dealing with complex problems, several aspects must be carefully considered to make enumeration viable.

One major challenge is the growth of the search space when dealing with complex instances. For a formula $F$ with $n$ variables, there are $2^n$ possible total assignments. Explicitly enumerating all of these solutions would require exponential space, which is impractical for large values of $n$. To mitigate this issue, \textbf{partial models} can be utilized to provide more concise representations of the solution set. A partial model is an assignment that leaves some variables unspecified, implying that the truth value of these variables does not influence the satisfiability of the formula in that particular assignment. Consequently, a partial assignment with $m$ specified variables represents $2^{n-m}$ total assignments, effectively reducing the solution space to explore.

Another critical aspect to consider in enumeration is the handling of repeated models. In some contexts, allowing the repetition of the same model in the enumeration might be acceptable or even desirable, such as in predicate abstraction applications. However, in other scenarios like Weighted Model Integration and \#SMT, repeating the same model can lead to incorrect results or inefficiencies. 
In this paper, we focus on disjoint enumeration, where repetitions of the same assignment are strictly prohibited.

\paragraph{Related work} SAT-based enumeration algorithms can be categorized into two main types: {\bf blocking solvers} and {\bf non-blocking solvers}.

Blocking AllSAT solvers \cite{mcmillan2002applying, jin2005efficient, yu2014all} are built on top of Conflict-Driven Clause Learning (CDCL) and non-chronological backtracking (NCB). These solvers work by adding {\bf blocking clauses} to the formula each time a model is found. A blocking clause is designed to exclude the current satisfying assignment, ensuring that the solver does not find and return the same assignment in subsequent searches. This process is repeated until all possible satisfying assignments have been enumerated, effectively scanning the entire search space. Even though blocking solvers are relatively straightforward to implement and can be modified to retrieve partial assignments, they face significant efficiency challenges as the number of models increases. Specifically, an exponential number of blocking clauses might be required to cover the search space entirely. As more blocking clauses are added, unit propagation—the process of deducing variable assignments from the existing clauses—becomes increasingly difficult, leading to degraded performance.

Non-blocking AllSAT solvers \cite{grumberg2004memory, li2004novel} address the inefficiencies associated with blocking clauses by avoiding their use altogether. Instead, these solvers employ chronological backtracking (CB) \cite{nadel2018chronological}. In chronological backtracking, when a conflict arises during the search process, the solver backtracks to the most recently assigned variable, rather than jumping back non-chronologically as in CDCL. This method avoids covering the same model multiple times without the performance degradation caused by an excessive number of blocking clauses.
However, non-blocking solvers have their limitations. They only generate total assignments, meaning all variables must be assigned a value in each satisfying assignment. Additionally, non-blocking solvers can struggle to efficiently escape regions of the search space that contain no solutions, which can lead to inefficiencies in certain scenarios.

\cite{mohle2019combining} introduces a formal calculus for disjunctive model counting that seeks to combine the strengths of both chronological backtracking and CDCL. This approach offers a promising direction, but the original work did not include an implementation or empirical results to demonstrate its effectiveness. Moreover, the calculus does not address how to effectively handle projected enumeration, and extending these methods to problems that include first-order logic theories.

\begin{gschange}
Another SAT-based approach for enumeration that is particularly useful for non-CNF formulas is based on the idea of {\it entailment} \cite{sebastiani2020you}. Typically, when given a partial assignment $\mu$, SAT solvers check whether $\mu$ satisfies $F$ by substituting all the assigned variables in $F$ with their corresponding truth values and recursively propagating these values through the formula. If this process results in $\top$, then $\mu$ is said to satisfy $F$, a concept known as "evaluation to true." Entailment, on the other hand, operates differently. A partial assignment $\mu$ entails $F$ if every total assignment $\eta$ that extends $\mu$ also satisfies $F$. In other words, after substituting the variables in $F$ with the values from $\mu$ and propagating, the residual formula must be valid. Whereas determining whether a partial assignment entails a formula is computationally more expensive than simply checking if the formula evaluates to true given $\mu$, it has been shown to be effective in generating compact partial models for enumeration \cite{fried2024entailing}. A few formal calculi have been developed to implement enumeration algorithms that leverage dual reasoning during enumeration \cite{mohle2020four, mohle2021enumerating}.

An alternative to SAT-based methods involves compiling a formula into a decomposable, deterministic negation normal form (d-DNNF). This structure allows for efficient retrieval of partial assignments that satisfy $F$. In a d-DNNF representation of a formula, every path from the root to a leaf node corresponds to a (possibly partial) assignment that satisfies $F$, and the properties of d-DNNF ensure that each assignment is mutually exclusive from the others. Building on this concept, a recently proposed tool leverages a depth-first traversal of the d-DNNF representation of a Boolean formula to enumerate models, ensuring that memory usage remains bounded by the size of $F$ \cite{lagniez2024leveraging}. It is important to note, however, that this approach is inherently designed for AllSAT and does not consider projection or first-order logic theories.
Recently a new calculus to convert CNF formulas to d-DNNF using chronological backtracking has been presented \cite{mohle2023abstract}.

The literature on AllSMT is very limited, and AllSMT algorithms are highly based on AllSAT techniques and tools. For instance, {\sc MathSAT5}~\cite{cimatti2013mathsat5} implements an AllSMT functionality based on the procedure by~\cite{lahiri2006smt}, relying on learning blocking clause to search all possible satisfying assignments.

\paragraph{Our Contribution} 

Based on the formal calculus in \cite{mohle2019combining}, in this paper we present \solverPlus{} and \solverSMT{}, respectively a projected AllSAT solver and a projected AllSMT solver that combine CDCL and chronological backtracking to avoid the introduction of blocking clauses. In particular, our main contributions can be summarized as follows:


    \begin{enumerate}
        \item[($a$)] We discuss the AllSAT
        procedure to perform disjoint partial enumeration of propositional
        formulas by combining the best of current All-
        SAT state-of-the-art literature: ($i$) CDCL, to escape search
        branches where no satisfiable assignments can be found; ($ii$)
        chronological backtracking, to ensure no blocking clauses
        are introduced; ($iii$) efficient implicant shrinking, to reduce
        in size partial assignments, by exploiting the 2-literal watching
        scheme.
        \item[($b$)] We propose two implicant shrinking algorithms, intending to reduce the number of partial assignments retrived by the AllSAT procedure while making sure the calculus in \cite{mohle2019combining} is not violated. 
        \item[($c$)] We extend our procedure to deal with projected enumeration, showing how chronological backtracking and CDCL have been adapted to enumerate only a subset of important variables.
        \item[($d$)] We extend our procedure to deal with the enumeration of SMT formulas, showing how chronological backtracking and CDCL have been adapted to integrate theory reasoning.
        \item[($e$)] We perform an extensive experimental evaluation, showing the superiority of our proposed techniques against the state-of-the-art algorithms.
    \end{enumerate}

\paragraph{Disclaimer} A preliminary and much shorter version of this paper was presented at the AAAI24 conference \cite{spallitta2024disjoint}, presenting only the baseline algorithm to perform disjoint enumeration without introducing blocking clauses ($a$) and the first of the two chronological implicant shrinking algorithm discussed in this work ($b)$. We refer to this baseline algorithm in the manuscript as \solver{}. 

This paper leverages the algorithm from \cite{spallitta2024disjoint} by proposing a novel implicant shrinking that allows the retrieval of shorter partial assignments without affecting computational times and is not heavily affected by variable ordering as the baseline algorithms ($b$). Moreover, we extended the algorithm to deal with projected enumeration ($c$) and SMT enumeration ($d$). Finally we provided an extensive and detailed experimental evaluation, to compare the implicant shrinking algorithms and the novel \solverPlus{} algorithm against the state-of-the-art solvers ($e$).

\paragraph{Structure of the paper} The rest of the paper is organized as follows. In~\sref{sec:preliminaries} we introduce the background, focusing on the notation adopted, the CDCL algorithm, and chronological backtracking. In~\sref{sec:solver} we briefly summarize the algorithm to perform AllSAT integrating CDCL and chronological backtracking with no need for blocking clauses, representing a summary of \cite{spallitta2024disjoint} and the baseline of extensions discussed in this work. In~\sref{sec:aggressive} we discuss the novel implicant shrinking algorithm, to make it more effective. In~\sref{sec:projected} and ~\sref{sec:SMT} we extend the baseline algorithm to address respectively projected SAT enumeration\cite{mohle2019combining} and SMT enumeration, showing the main difference and design choices that were required to make it compliant with the calculus in \cite{mohle2019combining}. At the end of sections \sref{sec:aggressive}, \sref{sec:projected} and \sref{sec:SMT} an extensive experimental evaluation is presented, comparing our tool's latest version against the few publicly available state-of-the-art competitors.

\end{gschange}

\section{Background}
\label{sec:preliminaries}

\subsection{Notation} 

 We assume $F$ is a propositional formula defined on the set of Boolean variables $V = \{v_1, ..., v_n\}$, with cardinality $|V|$. A \textit{literal} $\ell$ is a variable $v$ or its negation $\neg v$. The function $var(\ell)$ maps a literal to the associated variable. When dealing with projected enumeration, the set of variables $V$ is split into two disjoint sets: the set of relevant variables $V_r$ and the set of irrelevant variables $V_i$. $L(V)$ denotes the set of literals on V. We implicitly remove double negations: if $\ell$ is $\neg v$, by $\neg \ell$ we mean $v$ rather than $\neg\neg v$. A \textit{clause} is the disjunction of literals $\bigvee_{\ell\in c} \ell$. A \textit{cube} is the conjunction of literals $\bigwedge_{\ell\in c} \ell$. 

 A function $M: V \mapsto \{\top, \perp \}$ mapping variables in $F$ to their truth value is known as \textit{assignment}. An assignment can be represented by either a set of literals $\{\ell_1, ..., \ell_n\}$ or a cube conjoining all literals in the assignment $\ell_1 \wedge ... \wedge \ell_n$.
 We distinguish between \textit{total assignments} $\eta$ or \textit{partial assignments} $\mu$ depending on whether all variables are mapped to a truth value or not, respectively.
 
 A \textit{trail} is an ordered sequence of literals $I = \ell_1, ..., \ell_n$ with no duplicate variables. The empty trail is represented by $\varepsilon$. Two trails can be conjoined one after the other $I = KL$, assuming $K$ and $L$ have no variables in common. We use superscripts to mark literals in a trail $I$: $\ell^d$ indicates a literal assigned during the decision phase, whereas $\ell^*$ refers to literals whose truth value is negated due to chronological backtracking after finding a model (we will refer to this action as \textit{flipping}). Trails can be seen as ordered \emph{total} (resp. \emph{partial}) assignments; for the sake
of simplicity, we will refer to them as \emph{total} (resp. \emph{partial}) trails.

 \begin{definition}
     The \emph{decision level function} $\delta: V \mapsto \mathbb{N} \cup \{\infty\}$ returns the decision level of variable $V$, where $\infty$ means unassigned. We extend this concept to literals ($\delta(\ell) = \delta(var(\ell))$) and clauses ($\delta(C) = \{max(\delta(\ell)) |\ell \in C\}$).
 \end{definition}

  \begin{definition}
     The \emph{decision literal function} $\sigma: \mathbb{N} \mapsto L(V) \cup \{\varepsilon\}$ returns the decision literal of a specified level. If we have not decided on a literal at the specified level yet, we return $\varepsilon$. 
 \end{definition}

\begin{definition}
    The \emph{reason function} $\rho(\ell)$ returns the reason that forced literal $\ell$ to be assigned a truth value:
\begin{itemize}
    \item \textsc{DECISION}, if the literal is assigned by the decision selection procedure;
    \item \textsc{UNIT}, if the literal is unit propagated at decision level 0, thus it is an initial literal;
    \item \textsc{PROPAGATED($c$)}, if the literal is unit propagated at a decision level higher than 0 due to clause $c$;
\end{itemize} 
\end{definition}

\begin{gschange}
\subsection{AllSAT, AllSMT and Projection}

AllSAT is the task of enumerating all the truth assignments 
propositionally 
satisfying a propositional formula.\@ 
The task can be found in the literature in two versions: \emph{disjoint} AllSAT, in which the assignments are required to be pairwise mutually inconsistent, and \emph{non-disjoint} AllSAT, in which they are not. In this paper, we will focus on the former.
A generalization to the \smtt{} case is All\smtt{}, defined as the
task of enumerating all the \T{}{\em -satisfiable} truth assignments propositionally satisfying a \smtt{} formula. 

Projection is a process related to SAT and AllSAT solving that involves ignoring irrelevant variables $V_i$ from a Boolean formula while preserving the satisfiability of the formula with respect to the remaining relevant variables $V_r$. The goal of projection is to reduce the dimensionality of a Boolean expression by ``projecting" it onto a subset of its variables, effectively discarding those that are not relevant to the problem at hand. In particular, given a formula $F$ under the set of variables $V_r \cup V_i$ s.t. $V_r \cap V_i = \emptyset$, the enumeration of $F$ projected onto the relevant variables $V_r$ consists of:

\begin{equation}
    ProjAllSAT(F(V_r, V_i)) = AllSAT(\exists V_i. F(V_r,V_i))
\end{equation}

For example, the enumeration of a non-CNF formula $F$ can be carried out by first converting it into CNF and then enumerating its satisfying assignments by means of \emph{Projected AllSAT}. Specifically, given a non-CNF formula $F(\allA)$, we can apply either the Tseitin \cite{tseitinComplexityDerivationPropositional1983} or Plaisted-Greenbaum \cite{plaistedStructurepreservingClauseForm1986} transformation to obtain $\vicnf(\allA\cup\allB)$, where $\allB$ represents the Boolean variables introduced by the transformation. Enumeration is then performed over the partial assignments to $\allA$ that can be extended to total truth assignments satisfying $\vicnf$ over $\allA\cup\allB$. Here, the original set of variables $\allA$ corresponds to $V_r$, whereas the additional variables $\allB$, introduced during the CNF transformation, correspond to $V_i$.
We refer the reader to \cite{masina2023cnf} for an analysis of CNF-ization for enumeration.

\end{gschange}

 \subsection{The 2-Watched Literal Scheme}
 \label{sec:2watched}

 The \emph{2-watched literal scheme} \cite{moskewicz2001chaff} is an
indexing technique that efficiently checks if the currently-assigned
literals do not cause a conflict. For every clause, two literals are tracked. If at least one of the two literals is set to $\top$, then the clause is satisfied. If one of the two literals is set to $\perp$, then we scan the clause searching for a new literal $\ell'$ that can be paired with the remaining one, being sure $\ell'$ is not mapped to $\perp$. If we reach the end of the clause and both watches for that clause are set to false, then we know the current assignment falsifies the formula. The 2-watched literal scheme is implemented through watch lists.

 \begin{definition}
The \emph{watch list function} $\omega(\ell)$ returns the set of clauses $\{c_1,...,c_n\}$ currently watched by literal $\ell$.
\end{definition}

 \subsection{CDCL and Non-chronological Backtracking}

 Conflict Driven Clause Learning (CDCL) is the most popular SAT-solving technique \cite{marques1999grasp}. It is an extension of the older Davis-Putnam-Logemann-Loveland (DPLL) algorithm \cite{davis1962machine}, improving the latter by dynamically learning new clauses during the search process and using them to drive backtracking.

Every time the current trail falsifies a formula $F$, the SAT solver generates a conflict clause $c$ starting from the falsified clause, by repeatedly resolving against the clauses which caused unit propagation of falsified literals. This clause is then learned by the solver and added to $F$. Depending on $c$, we backtrack to flip the value of one literal, potentially jumping more than one decision level (thus we talk about \emph{non-chronological backtracking}, or NBC). CDCL and non-chronological backtracking allow for escaping regions of the search space where no satisfying assignments are admitted, which benefits both SAT and AllSAT solving.
The idea behind conflict clauses has been extended in AllSAT to learn clauses from partial satisfying assignments (known in the literature as \emph{good learning} or \emph{blocking clauses} \cite{bayardo2000counting,morgado2005good}) to ensure no total assignment is covered twice.

 \subsection{Chronological Backtracking}

\GSCHANGEBIS{Chronological backtracking (CB) is the core of the original DPLL algorithm. Considered inefficient for SAT solving once NBC was presented in \cite{moskewicz2001chaff}, it was recently revamped for both SAT and AllSAT in \cite{nadel2018chronological, mohle2019backing}.}
The intuition is that non-chronological backtracking after conflict analysis can lead to redundant work, due to some assignments that could be repeated later on during the search.
Instead, independently of the generated conflict clause $c$ we chronologically backtrack and flip the last decision literal in the trail. Consequently, we explore the search space systematically and efficiently, ensuring no assignment is covered twice during execution. Chronological backtracking combined with CDCL is effective in SAT solving when dealing with satisfiable instances. In AllSAT solving, it guarantees blocking clauses are no more needed to ensure termination.

\begin{gschange}
\subsection{Benefits and Drawbacks of CDCL and CB for Enumeration}

Enumeration is generally more challenging compared to SAT solving. In SAT solving, the search terminates as soon as a solution is found, whereas enumeration requires exploring the entire search space to identify all possible solutions. This makes the task of enumeration strictly more difficult than finding a single satisfying assignment. 

Considering SAT-based enumeration algorithms, there is no clear supremacy between blocking and non-blocking AllSAT solvers.
In particular, we can highlight the following strengths and weaknesses:

\begin{itemize}
    \item \textbf{Systematic Search:} chronological backtracking systematically scans the entire search space, ensuring that all regions are visited without repetition, particularly regions with no solution. CDCL, on the other hand, may not guarantee that some areas are not visited more than once, revisiting areas with no solution multiple times, unless blocking clauses are added.
    \item \textbf{Blocking clauses:} due to the systematic nature of chronological backtracking, there is no need for additional blocking clauses to prevent redundant exploration of the search space. In contrast, CDCL relies on blocking clauses to avoid revisiting previously explored areas. This may require adding an up-to exponential number of blocking clauses,
    causing memory blowups and a degradation of unit propagation performances.
    \item \textbf{Conflict Analysis:} in areas of the search space with no solution, CDCL can leverage conflict analysis to escape and redirect the search to other regions quickly. Chronological backtracking, on the other hand, may become trapped in such regions until the entire sub-search space is fully explored.
    \item \textbf{Time efficiency:} due to its ability to escape regions of the search space with no solution, CDCL-based approaches can generally enumerate solutions faster than algorithms based on chronological backtracking.
    \item \textbf{Shrinking Techniques and Partial Assignments:} whereas there is extensive discussion in the literature regarding shrinking techniques associated with CDCL for enumeration and the generation of partial assignments, so far these topics have not been addressed in the context of chronological backtracking.
\end{itemize}
\end{gschange}

\section{Enumerating Disjoint Partial Models without Blocking Clauses}
\label{sec:solver}

 We summarize the approach allowing for enumerating disjoint partial models with no need for blocking clauses discussed in \cite{spallitta2024disjoint}, that integrates: Conflict-Driven Clause-Learning ({\bf CDCL}), to escape search branches where no satisfiable assignments can be found;
 Chronological Backtracking ({\bf CB}), to ensure no blocking clauses are introduced; and methods for shrinking models ({\bf Implicant Shrinking}),
 to reduce in size partial assignments.

 \begin{gschange}
     Several algorithms are proposed in this section, and we use a colored notation to mark significant differences with respect to baseline AllSAT solving and extensions to the original algorithm presented in \cite{spallitta2024disjoint}. In particular:

     \begin{itemize}
         \item For Algorithms 1-3, we color in \textcolor{red}{{\bf red}} all lines that differ from the baseline CDCL AllSAT solving algorithm.
         \item For all algorithms, we color in \textcolor{darkgreen}{{\bf green}} additional conditions and procedures that must be executed to perform projected enumeration.
         \item For all algorithms, we color in \textcolor{blue}{{\bf blue}} additional conditions and procedures that must be executed to perform SMT-based enumeration.
     \end{itemize}
 \end{gschange}

\subsection{Disjoint AllSAT by Integrating CDCL and CB}
\label{sec:solver-search}

The work in \cite{mohle2019combining} exclusively describes the calculus and a formal proof of correctness for a model counting
framework on top of CDCL and CB,  with neither any algorithm nor any reference in modern state-of-the-art solvers. To this extent, we
start by presenting an AllSAT procedure for the search algorithm combining the two techniques, which are reported in this section. \GMCHANGE{In particular, we highlight the major differences to a classical AllSAT solver implemented on top of CDCL and NBC}.

\GMCHANGE{Algorithm \ref{algo:chronocdcl} presents the main search loop of the AllSAT algorithm. 
(In Alg. 1 the reader is supposed to ignore the \blue{blue lines from 8 to 11}, which refer to the SMT version of the algorithm and which will be illustrated in \S6.)}

The goal is to find a total trail \trail{} that satisfies $F$. At each decision level, it iteratively decides one of the unassigned variables in $F$ and assigns a truth value (Alg. \ref{algo:chronocdcl}, lines \ref{algo:decide}-\ref{algo:decide-end}); it then performs unit propagation
(Alg. \ref{algo:chronocdcl}, line \ref{algo:unit}) until either a conflict is reached (Alg. \ref{algo:chronocdcl}, lines \ref{algo:conflict}-\ref{algo:conflict-end}), or no other variable can be
unit propagated leading to a satisfying total assignment (Alg. \ref{algo:chronocdcl}, lines \ref{algo:noconflict}-\ref{algo:noconflict-end}) or \textsc{Decide} has to be called again (Alg. \ref{algo:chronocdcl}, lines \ref{algo:decide}-\ref{algo:decide-end}).

\begin{algorithm}[t]
\begin{algorithmic}[1]
  \caption[A]{{\sc Chrono-CDCL}($F, V$)}%
  \label{algo:chronocdcl}
  \STATE $T \leftarrow \varepsilon$
  \STATE $dl \leftarrow 0$
  \WHILE{\textbf{true}}
  \STATE $T, c \leftarrow$ \textsc{UnitPropagation()} \label{algo:unit}
  \IF{$c \neq \varepsilon$}\label{algo:conflict} 
    \STATE \textsc{AnalyzeConflict}($T,c, dl$) \label{algo:conflict-stop}
    \ELSIF{$|T| = |V|$} \label{algo:noconflict}
        \STATE \textcolor{blue}{$c_T \leftarrow $ {\sc Check-Theory-Consistency}($T$)} \label{alg:tcheck}
        \IF{\textcolor{blue}{$c_T \neq \varepsilon$}} \label{alg:tcheck-bad}
            \STATE \textcolor{blue}{\textsc{AnalyzeConflict}($T,c_T, dl$)}
            \STATE \textcolor{blue}{\textbf{continue}} \label{alg:tcheck-bad-end}
        \ENDIF
        \STATE \textsc{AnalyzeAssignment($T, dl$)}\label{algo:noconflict-end}
  \ELSE
    \STATE \textsc{Decide}($T$) \label{algo:decide}
    \STATE $dl \leftarrow dl + 1$ \label{algo:decide-end}
  \ENDIF
  \ENDWHILE
\end{algorithmic}
\end{algorithm}

\GMCHANGE{Notice that the main loop is identical to an AllSAT solver based on non-chronological CDCL; the only differences are embedded in the procedure to get the conflict and the partial assignments. (We remark that from now on we color in \textcolor{red}{\textbf{red}} the lines that differ from the baseline CDCL AllSAT solver.)}

\begin{algorithm}[t]
\begin{algorithmic}[1]
  \caption[A]{{\sc AnalyzeConflict}($T, c, dl$)} \label{algo:get-conflict-full}
  \IF{\textcolor{red}{$\delta(c) < dl$}} \label{algo:backfirst}
        \STATE \textcolor{red}{$T \leftarrow$ \textsc{Backtrack($\delta(c)$)}} \label{algo:backfirst-end}
    \ENDIF
    \IF{\textcolor{black}{$dl = 0$}} \label{algo:terminate}
        \STATE \textcolor{black}{\textbf{terminate with all models found}} \label{algo:terminate-end}
    \ENDIF
    \STATE $\langle uip, c' \rangle \leftarrow$ \textsc{LastUIP-Analysis()} \label{algo:conflictanalysis}
    \STATE \textcolor{red}{$T \leftarrow$ \textsc{Backtrack($dl-1$)}} \label{algo:conflict-start}
    \STATE $T.push(\neg uip)$
    \STATE \textcolor{red}{$limit \leftarrow dl - 1$} \label{alg:safe-conf}
    \STATE $\rho(\neg uip) \leftarrow$ \textsc{Propagated($c'$)} \label{algo:conflict-end}
\end{algorithmic}
\end{algorithm}

\GMCHANGE{Suppose {\sc UnitPropagation} finds a conflict, returning one clause $c$ in $F$ which is falsified by the current trail \trail{}, so that we invoke {\sc AnalyzeConflict}. Algorithm \ref{algo:get-conflict-full} shows the procedure to either generate the conflict clause or stop the search for new assignments if all models have been found.} 

We first compute the maximum assignment level of all literals in the conflicting clause $c$ and
    backtrack to that decision level (Alg. \ref{algo:get-conflict-full}, lines \ref{algo:backfirst}-\ref{algo:backfirst-end}) if strictly smaller than $dl$. This \GMCHANGE{additional step, not contemplated by AllSAT solvers that use NCB,} is necessary to support out-of-order
    assignments, the core insight in chronological backtracking when integrated into CDCL as described in \cite{nadel2018chronological}. 
    
    \GMCHANGE{Apart from this first step, Alg. \ref{algo:get-conflict-full} behaves similarly to a standard conflict analysis algorithm.} If the solver
    reaches decision level 0 at this point, it means there are no more variables to flip and the whole search space has been visited,
    and we can terminate the algorithm (Alg. \ref{algo:get-conflict-full}, lines \ref{algo:terminate}-\ref{algo:terminate-end}). Otherwise, we perform conflict analysis up to the last Unique Implication Point (last UIP\GSCHANGEBIS{, i.e. the decision variable at the current decision level}), retrieving the conflict clause $c'$ (Alg. \ref{algo:get-conflict-full}, line \ref{algo:conflictanalysis}),
    as proposed in~\cite{mohle2019combining}. Finally, we perform backtracking \GMCHANGE{(notice how we force chronological backtracking independently from the decision level of the conflict clause)}, push the flipped UIP into the trail, and set $c'$ as its assignment reason for the flipping (Alg. \ref{algo:get-conflict-full}, lines \ref{algo:conflict-start}-\ref{algo:conflict-end}). (The reader should temporarily skip line 10: the role of variable $limit$ is explained in \sref{sec:new-shrink}).

    \begin{algorithm}[t]
\begin{algorithmic}[1]
  \caption[A]{{\sc AnalyzeAssignment}($T, dl$)} \label{algo:get-assignment-full}
  \STATE \textcolor{red}{$dl' \leftarrow $ \textsc{Implicant-Shrinking($T$)}} \label{algo:implishrink}
        \IF{\textcolor{red}{$dl' < dl$}} \label{algo:backshrink}
            \STATE \textcolor{red}{$T \leftarrow$ \textsc{Backtrack($dl'$)}} \label{algo:backshrink-end}
        \ENDIF
        \STATE \textbf{store model} $T$ \label{algo:print}
        \IF{$dl' = 0$} \label{algo:exit-true}
            \STATE \textbf{terminate with all models found} \label{algo:exit-true-end}
        \ELSE
            \STATE $\ell_{flip} \leftarrow \neg (\sigma(dl'))$ \label{algo:chronoback}
            \STATE \textcolor{red}{$T \leftarrow$ \textsc{Backtrack($dl' - 1$)}}
            \STATE $T.push(\ell_{flip})$
            \STATE \textcolor{red}{$limit \leftarrow dl' - 1$} \label{ref:safe-ass}
            \STATE $\rho(\ell_{flip}) =$ \textsc{Backtrue} \label{algo:chronoback-end}
        \ENDIF
    
\end{algorithmic}
\end{algorithm}

Suppose instead that every variable is assigned a truth value without generating conflicts (Alg. \ref{algo:chronocdcl}, line \ref{algo:noconflict}); \GMCHANGE{then 
the current total trail \trail{} satisfies $F$, and we invoke {\sc AnalyzeAssignment}. Algorithm \ref{algo:get-assignment-full} shows the steps to possibly shrink the assignment, store it and continue the search.}

First, \textsc{Implicant-Shrinking} checks if, for some decision level $dl'$, we can backtrack up to $dl' < dl$ and obtain a partial trail still satisfying the formula (Alg. \ref{algo:get-assignment-full}, lines \ref{algo:implishrink}-\ref{algo:backshrink-end}). 
(We discuss the details of chronological implicant shrinking in the next subsection.)
We can produce the current assignment from the current trail \trail{} (Alg. \ref{algo:get-assignment-full}, line \ref{algo:print}). Then we check if all variables in \trail{} are assigned at decision level $0$.
If this is the case, then this means that we found the last assignment to cover $F$, so that we can end the search (Alg. \ref{algo:get-assignment-full}, lines \ref{algo:exit-true}-\ref{algo:exit-true-end}). Otherwise, we perform chronological backtracking, flipping the truth value of the currently highest decision variables and searching for a new total trail \trail{} satisfying $F$ (Alg. \ref{algo:get-assignment-full}, lines \ref{algo:chronoback}-\ref{algo:chronoback-end}).

We remark that in
\cite{mohle2019combining} it is implicitly assumed that one can determine if a partial trail satisfies the formula right after being generated, whereas modern SAT
solvers cannot check this fact efficiently, and
detect satisfaction only when trails are total. To cope with this issue, in our approach the
partial trail satisfying the formula is computed {\em a posteriori} from the total one by implicant shrinking. \GSCHANGEBIS{Moreover, the mutual exclusivity among different assignments is guaranteed, since the shrinking of the assignments is performed so that the generated partial assignments fall under the conditions of \sref{sec:solver} in \cite{mohle2019combining}).}   

Notice that the calculus discussed in \cite{mohle2019combining} assumes the last UIP is the termination criteria for the conflict analysis. We provide the following counter-example to
show that the first UIP does not guarantee mutual exclusivity between returned assignments.

\begin{example}
\label{ex:fuzz}
    Let $F$ be the propositional formula:
    $$ F = \overbrace{(x_1 \vee \neg x_2)}^{c_1} \wedge \overbrace{(x_1 \vee \neg x_3)}^{c_2} \wedge \overbrace{(\neg x_1 \vee \neg x_2)}^{c_3} $$
    For the sake of simplicity, we assume \textsc{Chrono-CDCL} to return total truth assignments. If the initial variable ordering
    is $x_3, x_2, x_1$ (all set to false) then the first two total and the third partial trails generated by Algorithm \ref{algo:chronocdcl} are: 
    \begin{equation*}
        T_1 = \textcolor{black}{\neg x_3^d \neg x_2^d \neg x_1^d};\ \ 
    T_2 = \textcolor{black}{\neg x_3^d \neg x_2^d x_1^*};\ \ 
    T_3 = \textcolor{black}{\neg x_3^d x_2^*}
    \end{equation*}
    Notice how $T_3$ leads to a falsifying assignment: $x_2$ forces $x_1$ due to $c_1$ and $\neg x_1$ due to $c_3$ at the same time. A
    conflict arises and we adopt the first UIP algorithm to stop conflict analysis. We identify $x_2$ as the first unique implication point (UIP) and construct the conflict clause $\neg x_2$. Since this is a unit clause, we force its negation $\neg x_2$ as an initial unit. We can now set $x_3$ and $x_1$ to $\perp$ and obtain a satisfying assignment. The resulting total trail $T = \neg x_3 \neg x_2 \neg x_1$ is covered \textbf{twice} during the search process.
    \exdone{}
    
\end{example}

\GMCHANGE{
We also emphasize that the incorporation of restarts in the search algorithm (or any method that implicitly exploits restarts, such as rephasing) is not feasible, as reported in \cite{mohle2019combining}.
}

\ignore{
\subsection{Chronological Implicant Shrinking}
}

\label{sec:partial}

\subsection{Implicit Solution Reasons}
\label{sec:extend-allsat}

Incorporating chronological backtracking into the AllSAT algorithm makes blocking clauses unnecessary. Upon discovering a model, we backtrack chronologically to the most recently assigned decision variable $\ell$ and flip its truth value, as if there were a reason clause $c$ - containing the negated decision literals of \trail{} - that forces the flip. These reason clauses $c$ are typically irrelevant to SAT solving and are not stored in the system. On the other hand, when CDCL is combined with chronological backtracking, \GMCHANGE{these clauses are required for conflict analysis}.

\begin{example}
\label{ex:storing}
    Let $F$ be the same formula from Example \ref{ex:fuzz}.
    We assume the first trail generated by Algorithm \ref{algo:chronocdcl} is $T_1 = \neg x_3^d \neg x_2^d \neg x_1^d$. 
 Algorithm \ref{algo:shrinking} can reduce $x_1$ since $\neg x_2$ suffices to satisfy both $c_1$ and $c_3$. (More details about the minimization procedure are discussed in the next section, and they are not relevant for this example). Consequently, we obtain the assignment $\mu_1 = \neg x_3 \wedge \neg x_2$, then flip $\neg x_2$ to $x_2$. The new trail $I_2 = \neg x_3^d x_2^*$ forces $x_1$ to be true due to  $c_1$; then $c_3$ would not be satisfiable anymore and cause the generation of a conflict. The last UIP is $x_3$, so that the reason clause $c'$ forcing $x_2$ to be flipped must be handled by the solver to compute the conflict clause.\exdone{}
\end{example}

To cope with this fact, a straightforward approach would be storing these clauses in memory with no update to the literal watching indexing; this approach would allow for $c$ to be called exclusively by the CDCL procedure without affecting variable propagation. If $F$ admits a large number of models, however, storing these clauses would negatively affect performances, so either we had to frequently call flushing procedures to remove inactive backtrack reason clauses, or we could risk going out of memory to store them.

To overcome the issue, we introduce the notion of \textit{virtual backtrack reason clauses}. When a literal $\ell$ is flipped after a satisfying assignment is found, its reason clause contains the negation of decision literals assigned at a level lower than $\delta(\ell)$ and $\ell$ itself. Consequently, we introduce an additional value, \textsc{Backtrue}, to the possible answers of the reason function $\rho$. This value is used to tag literals flipped after a (possibly partial) assignment is found. When the conflict analysis algorithm encounters a literal $\ell$ having $\rho(\ell) = $ \textsc{Backtrue}, \GMCHANGE{the resolvent can be easily reconstructed by collecting all the decision literals with a lower level than $\ell$ and negating them. This way we do not need to explicitly store these clauses for conflict analysis, allowing us to save time and memory for clause flushing}.

\begin{gschange}
    It is important to note that an implicant shrinking algorithm cannot remove literals marked with a {\sc Backtrue} flag, as these are essential for ensuring that subsequent assignments remain mutually exclusive from previous ones. Specifically, for each literal with a {\sc Backtrue} reason, there exists an implicit blocking clause $C_b$ that includes $\ell$ and the negation of all decision literals up to $\ell$. While these blocking clauses are not explicitly generated, the implicant shrinking algorithm must ensure that no literal flagged with {\sc Backtrue} is dropped. Failing to preserve these literals would break the implicit blocking clauses, thereby compromising the disjointness of the assignments. With a similar reasoning, the first literal unit-propagated after conflict analysis cannot be removed from a trail, since it guarantees that the search space is sistematically scanned without repetitions. These remarks are fundamental when dealing with assignment shrinking and are further discussed in \sref{sec:new-shrink}.
\end{gschange}

\subsection{Decision Variable Ordering}
\label{sec:ordering}

As shown in \cite{mohle2019combining}, different orders during \textsc{Decide} can lead to a different number of partial trails retrieved if chronological backtracking is enabled. 
After an empirical evaluation, we set {\sc Decide} to select the priority score of a variable depending on the following ordered set of rules.

First, we rely on the Variable State Aware Decaying Sum \textit{(VSADS)} heuristic \cite{huang2005using} and set the priority of a variable according to two weighted factors: ($i$) the count of variable occurrences in the formula, as in the Dynamic Largest Combined Sum (DLCS) heuristics; and ($ii$) an ``activity score", which increases when the variable appears in conflict clauses and decreases otherwise, as in the Variable State Independent Decaying Sum (VSIDS) heuristic. If two variables have the same score, we set a higher priority to variables whose watch list is not empty. If there is still a tie, we rely on the lexicographic order of the names of the variables.
    
\section{Chronological Implicant Shrinking}
\label{sec:aggressive}

Effectively shrinking a total trail \trail{} when chronological backtracking is enabled is not trivial.

In principle, we could add a flag for each clause $c$ stating if $c$ is currently satisfied by the partial assignment or not, and check the status of all flags iteratively adding literals to the trail. Despite being easy to integrate into an AllSAT solver and avoiding assigning all variables a truth value, this approach is unfeasible in practice: every time a new literal $\ell$ is added/removed from the trail, we should check and eventually update the value of the flags of clauses containing it. In the long term, this would negatively affect performances, particularly when the formula has a large number of models.

Also, relying on implicant shrinking algorithms from the literature for NCB-based AllSAT solvers does not work for chronological backtracking. Prime-implicant shrinking algorithms do not guarantee the mutual exclusivity between different assignments, so that they are not useful in the context of disjoint AllSAT. Other assignment-shrinking algorithms, as in \cite{toda2016implementing}, work under the assumption that a blocking clause is introduced. 
For instance, suppose we perform disjoint AllSAT on the formula $F = x_1 \vee x_2$ and the ordered trail is $T_1 = x_1^d x_2^d$. A general assignment shrinking algorithm could retrieve the partial assignment $\mu = x_2$ satisfying $F$, but obtaining it by using chronological backtracking is not possible (it would require us to remove $x_1$ from the trail despite being assigned at a lower decision level than $x_2$) unless blocking clauses are introduced. 

In this context, we need an implicant shrinking algorithm such that: ($i$) it is compatible with chronological backtracking, i.e. we remove variables assigned at level $dl$ or higher as if they have never been assigned; ($ii$) it tries to cut the highest amount of literals while still ensuring mutual exclusivity.

\subsection{Chronological Implicant Shrinking based on 2-watched Literals}
\label{sec:old-shrink}

Considering all the aforementioned issues, \cite{spallitta2024disjoint} proposed a {\it chronological implicant shrinking} algorithm that used state-of-the-art SAT solver data structures (thus without requiring dual encoding), which is reported in Algorithm \ref{algo:shrinking}.

\begin{algorithm}[t!]
\begin{algorithmic}[1]
  \caption[A]{{\sc Implicant-Shrinking}($T$)}%
  \label{algo:shrinking}
  \STATE $b \leftarrow 0$ 
  \STATE $T' \leftarrow T$
  \WHILE{$T' \neq \varepsilon$} \label{algo:while-lift}
      \STATE $\ell \leftarrow T'.pop()$ \label{algo:etapop}
      \IF{$\rho(\ell) \neq $ \textsc{DECISION}} \label{algo:decisionnot}
            \STATE $b \leftarrow max(b, \delta(\ell))$ \label{algo:decisionnot-lineend}
                
    \ELSIF{$\delta(\ell) > b$}\label{algo:compare}
        \STATE $b \leftarrow$ \textsc{Check-Literal}$(\ell, b, T')$ \label{algo:simp}
    \ELSIF{$\delta(\ell) = 0$ or ($\delta(\ell) = b$ and $\rho(\ell) =$  \textsc{Decision})} \label{alg:shrinking-ending}
        \STATE \textbf{break} \label{alg:shrinking-ending-end}
    \ENDIF
       
  \ENDWHILE
  \STATE \textbf{return} $b$
\end{algorithmic}
\end{algorithm}

The idea is to pick literals from the current trail starting from the latest assigned literals (Alg. \ref{algo:shrinking}, lines \ref{algo:while-lift}-\ref{algo:etapop}) and determine the lowest decision level $b$ to backtrack and shrink the implicant. First, we check if $\ell$ was not assigned by \textsc{Decide} (Alg. \ref{algo:shrinking}, line \ref{algo:decisionnot}). If this is the case, we set $b$ to be at least as high as the decision level of $\ell$ ($\delta(\ell)$), ensuring that it will not be dropped by implicant shrinking (Alg. \ref{algo:shrinking}, line \ref{algo:decisionnot-lineend}), since $\ell$ has a role in performing disjoint AllSAT.

If this is not the case, we compare its decision level $\delta(\ell)$ to $b$ (Alg. \ref{algo:shrinking}, line \ref{algo:compare}). If $\delta(\ell) > b$, then
we actively check if it is necessary for \trail{} to satisfy $F$ (Alg. \ref{algo:shrinking}, line \ref{algo:simp}) and set $b$ accordingly. If \simplify{} tries to remove $\ell$ from the trail, we check for each clause $c$ watched by $\ell$ if the other 2-watched literal $\ell_2$ is in $T$ to determine if $\ell$ is necessary for the satisfiability of $F$, as if the clause $c$ is projected into the binary clause $\ell \vee \ell_2$. If $\ell_2$ is not in $T$, then we force the AllSAT solver to maintain $\ell$, setting the backtracking level to at least $\delta(\ell)$; otherwise we move on to the next clause watched by it. 

If $\ell$ is either an initial literal (i.e. assigned at decision level 0) or both $\rho(\ell) =$ \textsc{Decision} and $\delta(\ell) = b$ hold, all literals in the trail assigned before $\ell$ would have a decision level lower or equal than $b$. This means that we can exit the loop early (Alg. \ref{algo:shrinking}, lines \ref{alg:shrinking-ending}-\ref{alg:shrinking-ending-end}), since scanning further the trail would be unnecessary.
Finally, if none of the above conditions holds, we can assume that $b$ is already greater than $\delta(\ell)$, and we can move on to the next literal in the trail.

\begin{gschange}

This variant of implicant shrinking is conservative when it comes to dropping literals from the trail. We do not consider the possibility of another literal $\ell'$ , currently not watching $c$, being in the current trail $T$, and having a lower decision level than the two literals watching $c$.

\begin{example} \label{ex:bad_minimize}
    Let $F$ be the formula 
    \[
    F = (x_1 \vee x_2) \wedge (x_3 \vee x_4)
    \] 
    According to the variable ordering heuristic presented in \sref{sec:ordering}, all variables have the same VSADS score and all of them watch at least one clause. Consequently, the variable ordering will be $x_1, x_2, x_3, x_4$. Assume that every decision variable is set to a positive polarity, obtaining the assignment \[\eta = x_1^d x_2^d x_3^d x_4^d\] Whereas $x_4$ would be removed from the assignment by the shrinking procedure ($x_3$ ensure that all clauses where $x_4$ appear are satisfied), $x_3$ could not be removed, and the procedure would stop with the partial assignment \[\mu = x_1^d x_2^d x_3^d.\] Notice that this assignment could be further reduced to $\mu' = x_1^d x_3^d$, but due to the calculus in \cite{mohle2019combining} and the order chosen by the solver, the implicant shrinking procedure is forced to stop the shrinking early on. 
\end{example}

\subsection{Simulating Optimal Decision Variable Ordering}
\label{sec:new-shrink}

The previous example highlights an important aspect: if the solver knew an optimal variable ordering preventing from assigning as many variables as possible, then it should postpone their assignment to the very end. Whereas this kind of prediction is not feasible, we can try to simulate it. Once a total assignment $\eta$ is obtained, we can separate the variables in $\eta$ into two disjoint sets; ($i$) the variables that are necessary to satisfy all clauses; and ($ii$) the remaining unnecessary variables. 
If the search algorithm first assigns all necessary variables before the non-necessary ones, then the literals following the last necessary literal in the trail could be dropped without affecting satisfiability, as they were never critical to the assignment. Essentially, we can remove all non-necessary literals from a trail $T$, regardless of their position or decision level, assuming their truth value assignments could be deferred to the end of the main search loop. It is important to note that the order of literals in a trail $T$ does not influence whether the assignment satisfies $F$; any permutation of $T$ will still satisfy $F$.

Starting from this idea, we present a novel chronological implicant shrinking algorithm, which focuses on performing a more effective shrinking, whose general schema is shown in Algorithm \ref{algo:shrinking-plus}.

\begin{algorithm}[t!]
\begin{algorithmic}[1]
  \caption[A]{{\sc Implicant-Shrinking-Aggressive}($T$)}%
  \label{algo:shrinking-plus}
  \STATE $T', S, W, N \leftarrow T, \{\}, \{\}, \{\}$
  \STATE $W, N \leftarrow $ {\sc Initialize}($W,N$)
  \STATE $S \leftarrow $ {\sc Get-Important-Literals}($W,N,S$)
  \STATE $T, dl \leftarrow$ {\sc Lift-Literals}($S, T, dl$)
  \STATE \textbf{return} $dl$
\end{algorithmic}
\end{algorithm}

\begin{algorithm}[t!]
\begin{algorithmic}[1]
\caption[A]{{\sc Initialize}($W,N$)} \label{alg:initialization}
\FOR{$c\in F$} \label{alg:initi}
      \FOR{$\ell\in c$}
        \IF{$\ell\in T$}
          \STATE $W[\ell] = W[\ell] + c$
          \STATE $N[c] = N[c] + 1$
        \ENDIF
      \ENDFOR
  \ENDFOR \label{alg:initi-end}
  \STATE \textbf{return} $W, N$
\end{algorithmic}
\end{algorithm}

\begin{algorithm}[t!]
\begin{algorithmic}[1]
\caption[A]{{\sc Get-Important-Literals}($W,N,S$)} \label{alg:lift-aggressive}
\WHILE{$T' \neq \varepsilon$} \label{algo:while-lift-agg}
      \STATE $\ell \leftarrow T'.pop()$ \label{algo:etapop-lift}
      \IF{$\delta(\ell) \leq limit$ \textcolor{darkgreen}{\textbf{or} $\ell\notin V_r$}} \label{alg:safe}
        \STATE \textbf{continue} \label{alg:safe-end}
      \ENDIF
      \STATE $required \leftarrow false$
      \FOR{$c\in W[\ell]$} \label{alg:watch-check}
        \IF{$N[c] == 1$} \label{alg:watch-count}
            \STATE $required = true$
        \ENDIF
      \ENDFOR
      \IF{$required$} \label{alg:store}
        \STATE $S.push(\ell)$ \label{alg:store-end}
      \ELSE
      \FOR{$c\in W[\ell]$} \label{alg:drop}
        \STATE $N[c] = N[c] - 1$ \label{alg:drop-end}
      \ENDFOR
      \ENDIF
  \ENDWHILE
  \STATE \textbf{return} $S$
\end{algorithmic}
\end{algorithm}

\begin{algorithm}[t!]
\begin{algorithmic}[1]
\caption[A]{{\sc Lift-Literals}($S,T,dl$)} \label{alg:reorder}
\STATE $T \leftarrow$ \textsc{Backtrack($limit$)} \label{alg:backtracksafe}
  \STATE $dl \leftarrow limit$
  \FOR{$\ell\in S$} \label{alg:finalassign}
  \IF{$\ell\in T$} \label{alg:extracheck}
    \STATE \textbf{continue}
  \ENDIF
   \STATE $\textsc{Assign($\ell$)}$
   \STATE $dl \leftarrow dl + 1$
   \STATE $T, c \leftarrow$ \textsc{UnitPropagation()} \label{alg:unitit}
  \ENDFOR \label{alg:finalassignend}
  \STATE \textbf{return} $T, dl$
\end{algorithmic}
\end{algorithm}

We begin by initializing several auxiliary data structures (Alg. \ref{alg:initialization}): a copy of the satisfying total trail $T'$, an empty ordered list $S$ to store the literals from $T$ that form the shrunk partial assignment, a map $W: \ell \mapsto {c_1, ..., c_n}$ that links each literal in $T$ to the set of clauses containing it, and a map $N: c \mapsto \mathbb{N}$ that tracks how many literals in each clause are present in $T$.

We then proceed to determine the set of literals that cannot be lifted from the assignment, being necessary to satisfy $F$ (Alg. \ref{alg:lift-aggressive}).
(In Alg. 7 and 8 the reader is supposed to ignore the parts ``\green{{\bf or} $l\not \in V_r$}" in \green{green}, which refer to the projected version of the algorithmm and which will be illustrated in \S5.)
Starting from the most recently assigned literals (Alg. \ref{alg:lift-aggressive}, line \ref{algo:while-lift-agg}), we evaluate each literal $\ell$ to determine whether there exists a clause that is exclusively watched by $\ell$, indicated by a counter $N[c]$ being equal to 1 (Alg. \ref{alg:lift-aggressive}, lines \ref{alg:watch-check}-\ref{alg:watch-count}). If such a clause exists, $\ell$ cannot be dropped from the assignment and is added to $S$ (Alg. \ref{alg:lift-aggressive}, lines \ref{alg:store}-\ref{alg:store-end}). Conversely, if no clause necessitates $\ell$ for the shrunk assignment, $\ell$ is removed from $T$, and the map $N$ is updated accordingly for each clause containing $\ell$ (Alg. \ref{alg:lift-aggressive}, lines \ref{alg:drop}-\ref{alg:drop-end}). 

The algorithm, as described so far, does not account for chronological backtracking and might remove literals whose truth assignment is fundamental to enforce the disjointness of all assignments (as we remarked at the end of \sref{sec:ordering}). Specifically, this issue does arise when minimizing either \textsc{BackTrue} literal or literal propagated by conflict analysis. To ensure this does not happen, we introduce an auxiliary variable, $limit$, which stores the lowest level up to which literals cannot be dropped (Alg. \ref{alg:lift-aggressive}, lines \ref{alg:safe}-\ref{alg:safe-end}). The $limit$ variable is updated during each conflict analysis (Alg. \ref{algo:get-conflict-full}, line \ref{alg:safe-conf}) and at the end of each implicant shrinking procedure when a literal is flipped due to a \textsc{BackTrue} reason  (Alg. 3, line \ref{ref:safe-ass}). 

Once $S$ contains a  set of literals satisfying $F$, we conclude the shrinking procedure by dropping the remaining literals to simulate the best-case decision ordering scenario (Alg. \ref{alg:reorder}), where these removed literals would have been assigned later on and then dropped.
We first backtrack to the decision level stored in $limit$ (Alg. \ref{alg:reorder}, line \ref{alg:backtracksafe}), to ensure the entire search space is scanned correctly. After that, we add the remaining literals in $S$ that are not yet part of the current trail by assigning them one by one (Alg. \ref{alg:reorder}, lines \ref{alg:finalassign}-\ref{alg:finalassignend}). Each time a literal is assigned, we perform unit propagation (Alg \ref{alg:reorder}, line \ref{alg:unitit}) to ensure that any unassigned literals in $S$ that can have their truth value determined by unit propagation are correctly handled.

\begin{example} \label{ex:aggressive}
Consider the formula 
\[
F = \overbrace{(x_1 \lor x_2)}^{c_1} \land \overbrace{(x_3 \lor x_4)}^{c_2},
\]
and let 
\[
\eta_1 = x_1^d x_2^d x_3^d x_4^d
\]
be the first total assignment that satisfies \(F\), which we want to minimize. Since no conflict occurred before generating this assignment, \(\text{limit}\) is set to 0, meaning all literals are candidates for lifting. We initialize all auxiliary data structures, with \(N[c_1]\) and \(N[c_2]\) both starting at 2.

Starting with the most recent literal, \(x_4\) can be lifted, as no clause is satisfied solely by it. It is not added to \(S\), and \(N[c_2]\) is updated to 1. Next, \(x_3\) is processed, but it cannot be lifted: \(N[c_2] = 1\), so \(x_3\) is the only literal satisfying \(c_2\), and it is added to \(S\). Similarly, \(x_2\) can be dropped, while \(x_1\) must remain in the partial assignment. The ideal partial assignment generated is 
\[
\mu_1 = x_1^d x_3^d.
\]
To achieve this, we backtrack to level 0, clearing the trail, and explicitly reconstruct the partial trail \(\mu_1\) by reassigning the necessary truth values. This trail is stored by the solver, and due to chronological backtracking, \(x_3\) is flipped. At this point, \(\text{limit}\) is updated to 1, as \(\neg x_3\) is assigned at level 1.

After the decision and propagation procedures, the new total trail 
\[
\eta_2 = x_1^d \neg x_3^* x_4 x_2^d
\]
satisfies the formula. Here, \(\text{limit}\) is set to 1, meaning that to ensure only mutually exclusive partial assignments are generated, all literals assigned at levels up to \(\text{limit}\) cannot be lifted (in this case, \(x_1\), \(\neg x_3\), and \(x_4\)). On the other hand, no clause containing \(x_2\) requires it to be satisfied, so it can be dropped. The second partial trail satisfying \(F\) is then 
\[
\mu_2 = x_1^d \neg x_3^* x_4.
\]
 
\end{example}

The new implicant shrinking algorithm does have the drawback of needing to build $W$ and $N$ for each satisfying total assignment, which can impact performance. However, reducing the number and length of partial assignments is often the goal in many applications. Depending on the number of assignments covering a formula, the significant reduction in partial assignments and the consequent pruning of a large portion of the search space outweigh the minor inefficiency introduced by the implicant shrinking process.

We remark how the novel chronological implicant shrinking does not guarantee that the shrunk partial assignment is minimal. Minimality of a partial assignment means that given a partial assignment $\mu$ obtained by the search procedure, there is no other literal that can be removed by $\mu$ so that $\mu$ still satisfies $F$ and it is pairwise mutually exclusive against all the assignments retrieved before it.

\begin{example}
\label{ex:non-minimal}
    Consider the formula:

    \[
\varphi = (x_1 \lor x_2 \lor x_3) \land (x_1 \lor x_2 \lor \neg x_3) \land (x_1 \lor \neg x_2 \lor x_3) \land (\neg x_1 \lor x_2 \lor x_3)
\]

We assume the search algorithm favors negative polarity the first time a variable is chosen. The first trail generated by the algorithm is $\mu_1=\neg A_1^d\neg A_2^d\neg A_3$, the first two variables being chosen by {\sc Decide} and the last one being unit propagated because of the second clause. The first clause, however, is now falsified, and thus a conflict arises, forcing $\neg A_2$ to be flipped. To preserve mutual exclusivity between assignments, $limit$ should be updated up to 1, avoiding dropping anything before $A_2$. Now the current trail $\mu_2 = \neg A_1^d A_2^*$ is forced to add $A_3$ to satisfy the third clause. The total trail $\eta_1 = \neg A_1^d A_2^* A_3$ satisfies the formula. The implicant shrinking algorithm is not able to drop any literal; notice, however, that $\neg A_1$ could be dropped by $\eta_1$ without altering the satisfiability and the mutual exclusivity of the assignment against the currently empty set of assignments. For this reason, the algorithm is not guaranteed to find a minimal assignment. 
\end{example}

One could wonder if the implicant shrinking procedure could be modified for non-disjoint enumeration. In particular, we could argue that if we allow the algorithm to eliminate literals before the \(\textit{limit}\) level, then we could get shorter assignments sharing some of the total assignments under it with other partial models. However, the following example shows that under this assumption the AllSAT search is not guaranteed to terminate.

\begin{example}
    Consider the same formula of Example \ref{ex:non-minimal}, but this time the algorithm favors positive polarity for the first choice. we assume literals whose assignment level is lower than $limit$ can be dropped from the trail.

The search algorithm initially generates the first satisfying assignment, $\eta_1 = x_1^dx_2^dx_3^d$. Applying the shrinking process reduces this assignment to $\mu_1 = A_1^dA_2^d$, as $A_3$ can be removed without affecting the satisfaction of any clause exclusively depending on it.

Following chronological backtracking, we reach the trail $x_1^d \neg x_2^*$, which leads to the second satisfying assignment, $\eta_2 = x_1^d \neg x_2^* x_3$. In this case, $x_3$ is necessary to satisfy the final clause, while $x_2$ becomes redundant and can be dropped. The corresponding partial assignment is reduced to $\mu_2 = x_1^dx_3^d$. Notably, due to the reordering of literals, $x_3$ is now a decision literal, as it is no longer forced by $x_1$ and $\neg x_2$.
Upon flipping $x_3$, the algorithm encounters the trail $x_1^d \neg x_3^*$, which extends to the satisfying trail $\eta_3 = x_1^d \neg x_3^* x_2$. However, upon reduction, we again obtain $\mu_1$.

At this point, the algorithm cycles between $\mu_1$ and $\mu_2$, without any escape mechanism. The absence of blocking clauses prevents $x_1$ from being flipped, resulting in a recurring pattern of these partial assignments.
\end{example}

\subsection{Experimental evaluation}

We have implemented the ideas discussed in this paper in our tool \solverPlus{}, whose source code benchmarks are available on Zenodo \cite{spallitta_2024_14197776}. An updated version of the source code is available at \url{https://github.com/giuspek/tabularAllSAT}. Experiments are performed on an Intel Xeon Gold 6238R @ 2.20GHz 28 Core machine with 128 GB of RAM, running Ubuntu Linux 22.04. Timeout has been set to 1200 seconds. The experiments performed are the following:

\begin{itemize}
    \item Ablation study to compare the chronological implicant shrinking algorithm in \sref{sec:old-shrink} against the novel one proposed in \sref{sec:new-shrink} (\sref{sec:ablation});
    \item AllSAT experimental evaluation (\sref{sec:exp-allsat});
\end{itemize}

\subsubsection{Comparing implicant shrinking algorithms}
\label{sec:ablation}

We start our experimental evaluation by comparing the two chronological implicant shrinking algorithms discussed respectively in \sref{sec:old-shrink} and \sref{sec:new-shrink}. We consider the following benchmark, most of them being used in \cite{spallitta2024disjoint}:

\begin{itemize}
\item \textit{Rnd3sat} contains 410 random 3-SAT problems with $n$ variables, $n\in[10,50]$. In SAT instances, the ratio of clauses to variables needed to achieve maximum hardness is about 4.26, but in AllSAT, it should be set to approximately 1.5 \cite{bayardo1997using}. For this reason, we chose not to use the instances uploaded to SATLIB and we created new random 3-SAT problems accordingly.
\item We also tested our algorithms over SATLIB benchmarks, specifically \textit{CBS} and \textit{BMS} \cite{singer2000backbone}. 
\end{itemize}

\begin{figure}[t!]
        \centering
        \begin{subfigure}[b]{0.45\textwidth}
            \centering
            \includegraphics[width=\textwidth]{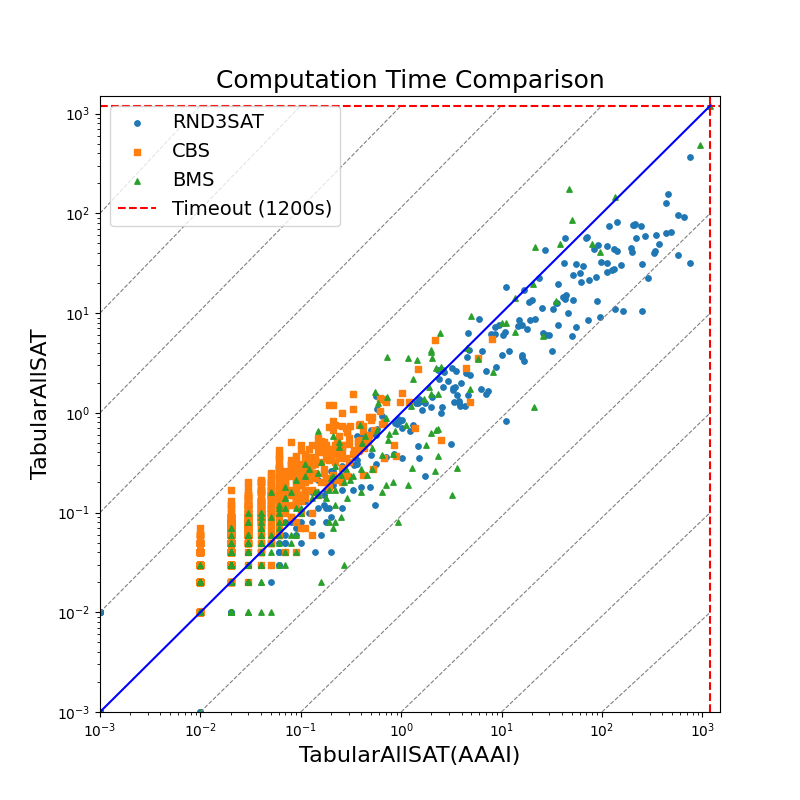}
            \caption%
            {{CPU Time (in seconds)}}    
            \label{fig:ablation-time}
        \end{subfigure}
        \begin{subfigure}[b]{0.45\textwidth}  
            \centering 
            \includegraphics[width=\textwidth]{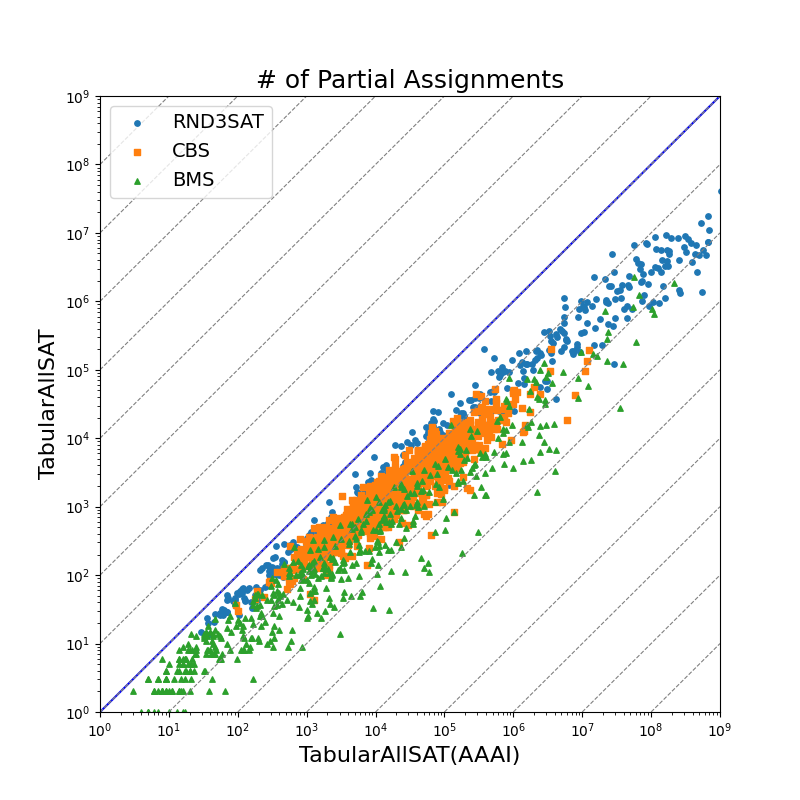}
            \caption[]%
            {{\# of partial models}}    
            \label{fig:albation-models}
        \end{subfigure}
        \caption{Scatter plot comparing CPU time and \# of partial models with the two implicant shrinking algorithms. The $x$ and $y$ axes are log-scaled.} 
        \label{fig:scatter-time}
    \end{figure}

We compared the two implicant shrinking algorithms on two metrics: ($i$) computational time, and ($ii$) number of partial assignments retrieved.
We checked the correctness of the enumeration by testing if the number of total assignments covered by the set of partial solutions was the same as the model count reported by the \#SAT solver Ganak \cite{SRSM19}, being always correct for both algorithms.
Figure \ref{fig:scatter-time} presents a log-scaled scatter plot comparison of two implicant shrinking algorithms, focusing on execution time (left) and the number of partial models generated (right). As expected, the novel implicant shrinking algorithm occasionally incurs a higher overhead due to the additional effort required to shrink assignments, which can result in slightly slower performance compared to the original algorithm with the easiest problems ($<$ 1s). However, the novel algorithm significantly reduces the number of partial assignments generated, with the impact becoming more pronounced as the number of total assignments for a given instance increases. In these larger instances, the novel implicant shrinking algorithm also demonstrates better performance in terms of execution time. All the following subsections' experiments rely on the novel implicant shrinking algorithm.

\subsubsection{Comparison against state-of-the-art solvers}
\label{sec:exp-allsat}

\begin{figure}[t!]
    \centering
    \begin{subfigure}[t]{0.32\textwidth}
        \centering
        \includegraphics[width=\textwidth]{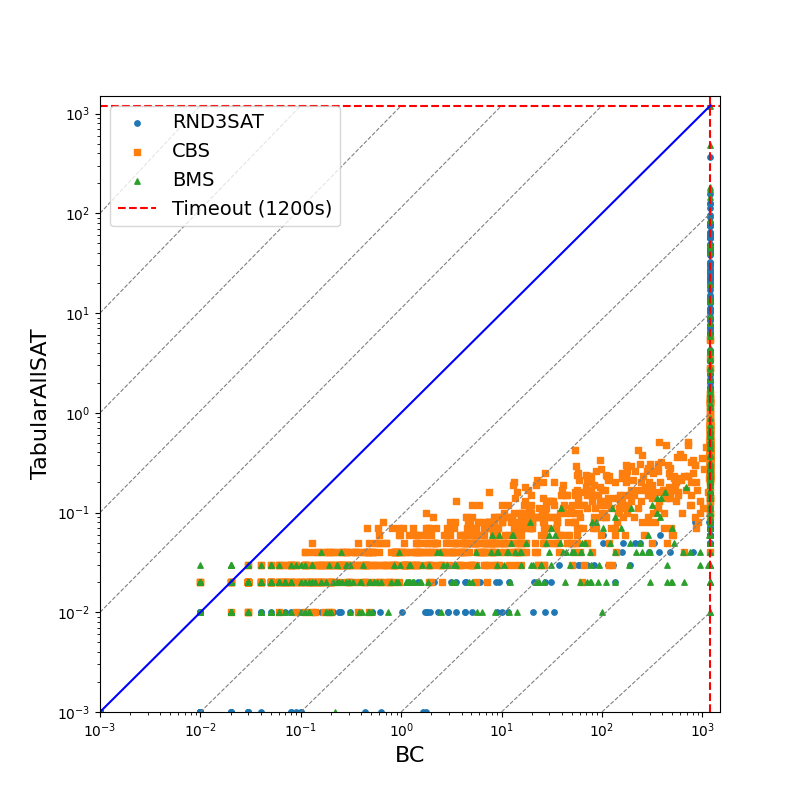}
        \caption{{\sc BC}}    
        \label{fig:soa-binary}
    \end{subfigure}
    \begin{subfigure}[t]{0.32\textwidth}  
        \centering 
        \includegraphics[width=\textwidth]{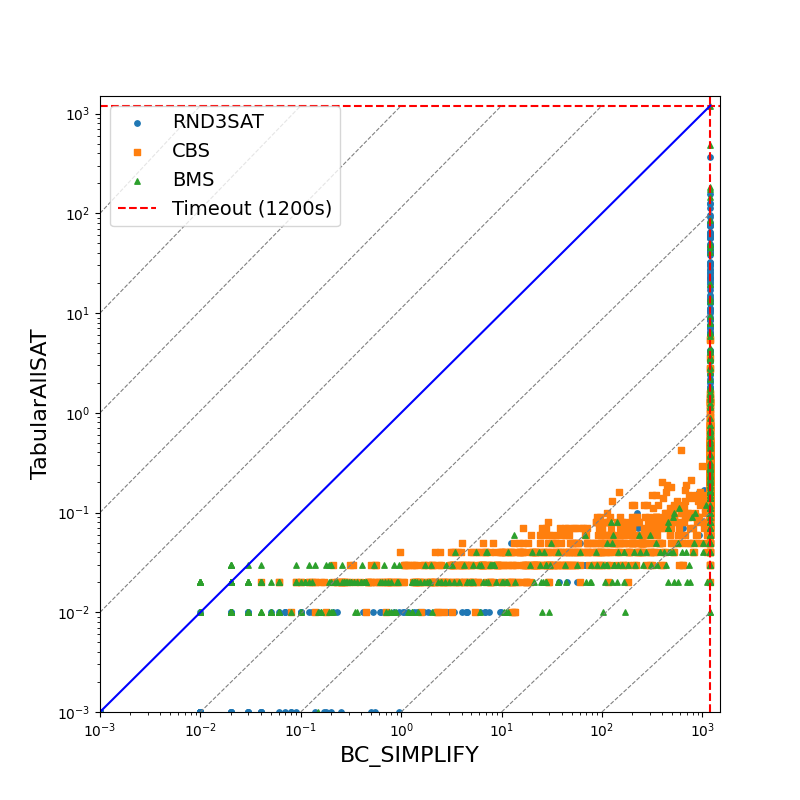}
        \caption{{\sc BC\_Partial}}    
        \label{fig:soa-CSB}
    \end{subfigure}
    \begin{subfigure}[t]{0.32\textwidth}   
        \centering 
        \includegraphics[width=\textwidth]{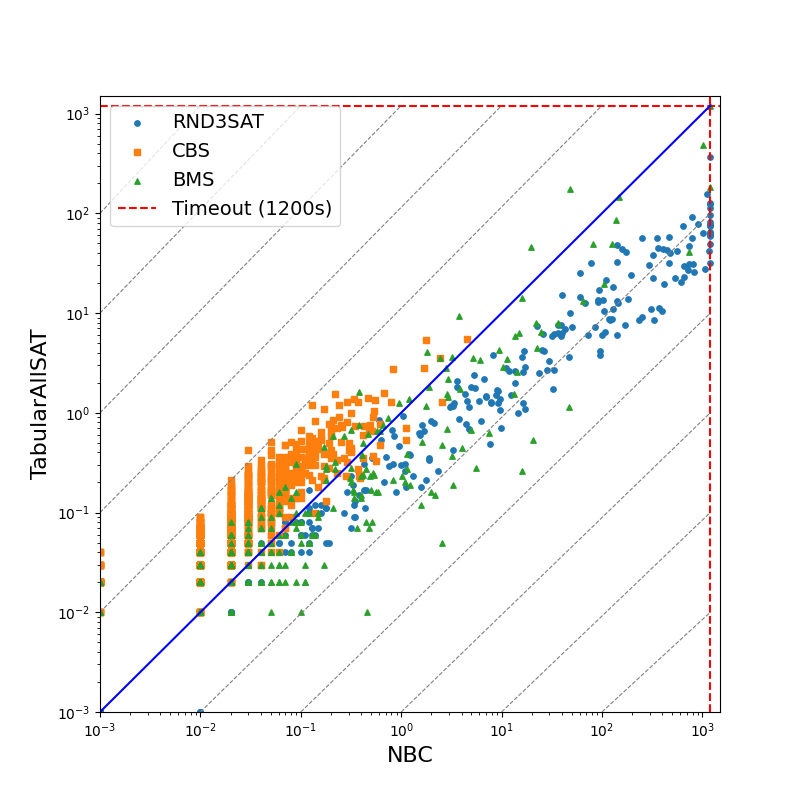}
        \caption{{\sc NBC}}    
        \label{fig:soa-rnd3sat}
    \end{subfigure}
    
    \begin{subfigure}[t]{0.32\textwidth}   
        \centering 
        \includegraphics[width=\textwidth]{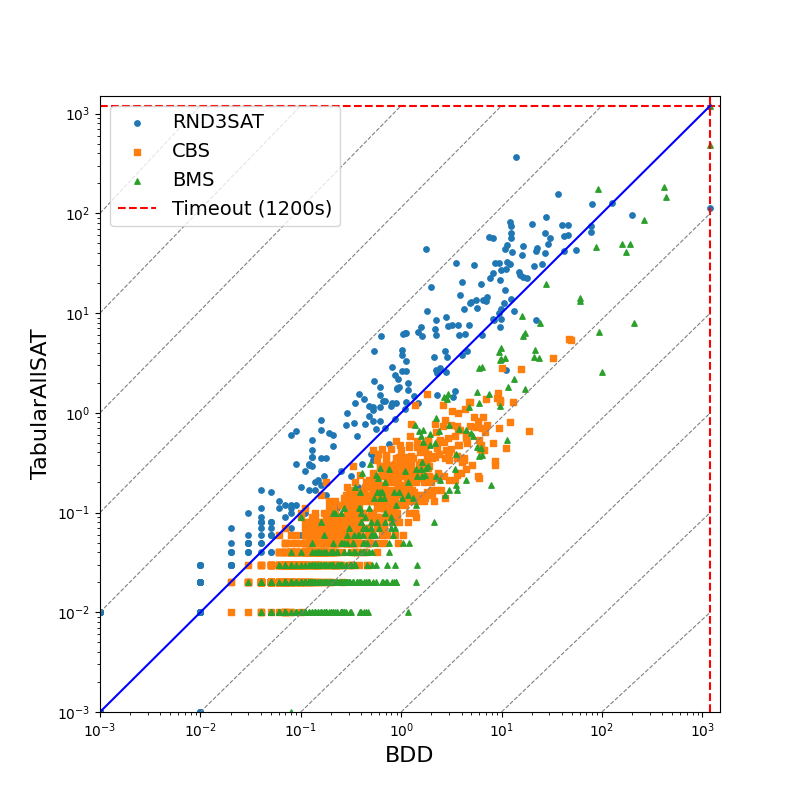}
        \caption{{\sc BDD}}    
        \label{fig:soa-BMSk3}
    \end{subfigure}
    \begin{subfigure}[t]{0.32\textwidth}   
        \centering 
        \includegraphics[width=\textwidth]{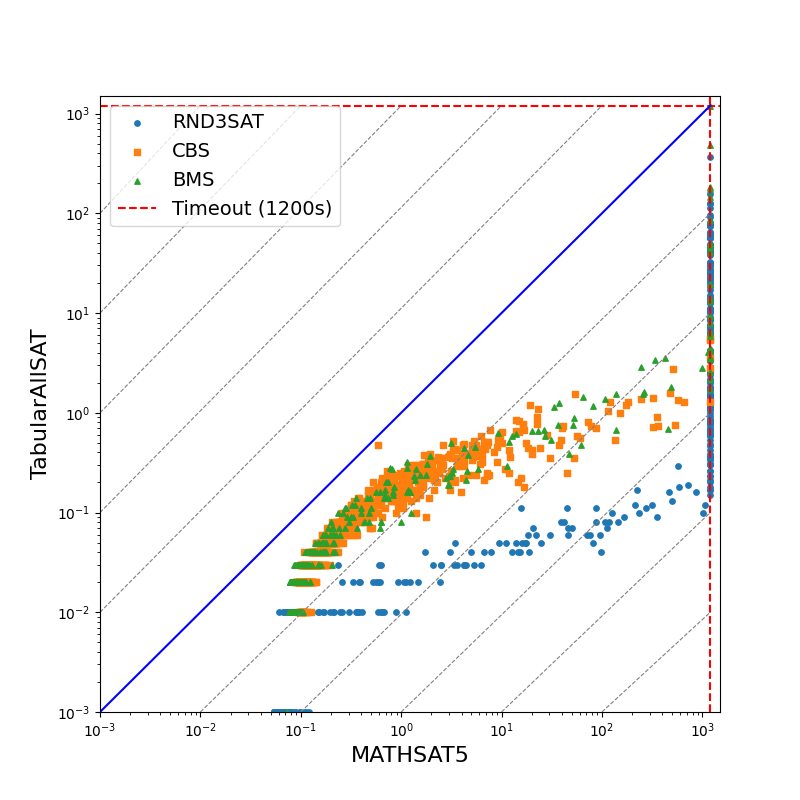}
        \caption{{\sc MathSAT5}}  
        \label{fig:sat-math-time}
    \end{subfigure}
    \begin{subfigure}[t]{0.32\textwidth}   
        \centering 
        \includegraphics[width=\textwidth]{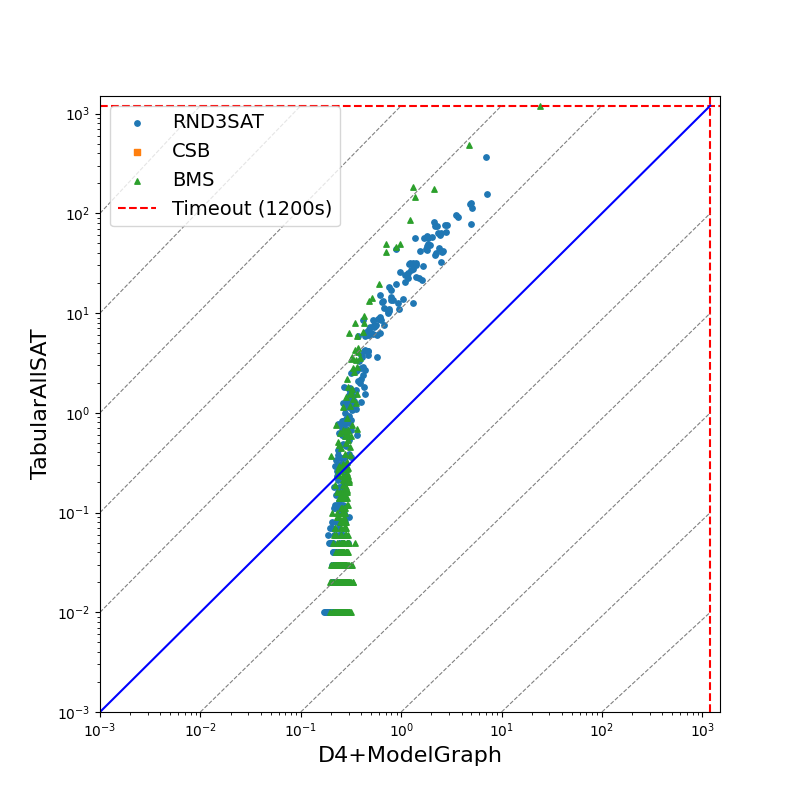}
        \caption{{\sc D4+ModelGraph}}    
        \label{fig:sat-kc-time}
    \end{subfigure}
    
    \caption{Scatter plots comparing \solverPlus{}
     CPU times against other AllSAT solvers. The $x$ and $y$ axes are log-scaled.} 
    \label{fig:stateofart}

    \vspace{0.5cm}

    \scriptsize
    \begin{tabular}{cccccccc}
                    & \solverPlus{} & {\sc D4+ModelGraph} & {\sc BDD} & {\sc NBC} & {\sc MathSAT5} & {\sc BC}   & {\sc BC\_Partial}     \\
        rnd3sat (410)       & \textbf{410} & \textbf{410}           & 409  & 396  & 229     & 194  & 210         \\
        CSB (1000)          & \textbf{1000} & \textbf{1000}         & \textbf{1000} & \textbf{1000} & 997     & 865  & 636         \\
        BMS (500)           & \textbf{500} & \textbf{500}           & 498  & 498  & 473     & 368  & 353         \\ \hline
        Total (1910)        & \textbf{1910} & \textbf{1910}          & 1907 & 1894 & 1699    & 1427 & 1199       
    \end{tabular}
    \caption{Number of instances solved by each solver within a timeout (1200 seconds) for the AllSAT benchmark.}
    \label{tb:table-timeout2}
\end{figure}

\begin{figure}
    \centering
    \begin{subfigure}[t]{0.32\textwidth}
        \centering
        \includegraphics[width=\textwidth]{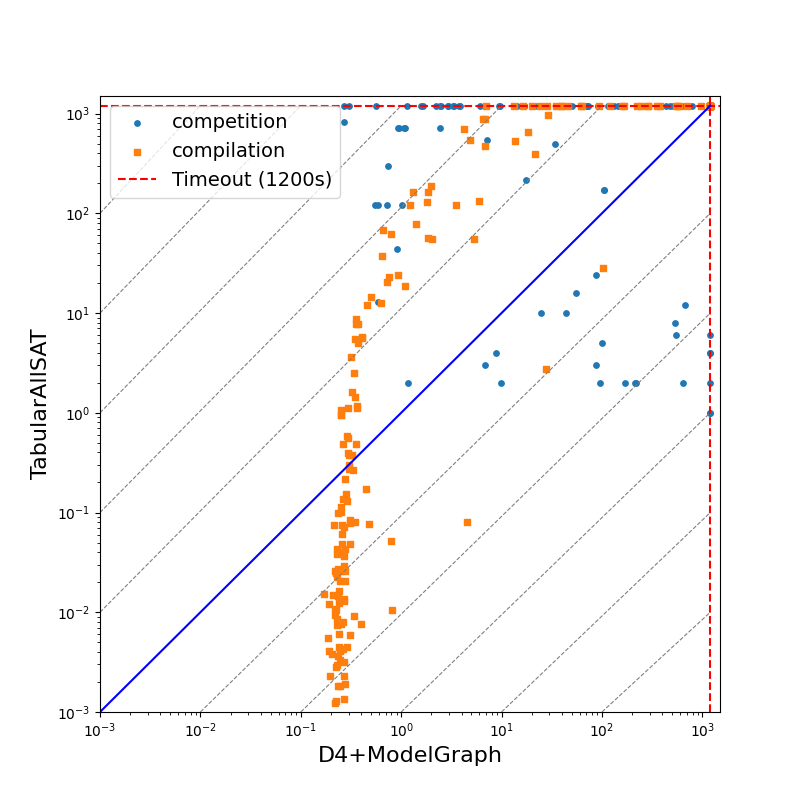}
        \caption{Time (in seconds)}    
        \label{fig:soa-hard-time}
    \end{subfigure}
    \begin{subfigure}[t]{0.32\textwidth}  
        \centering 
        \includegraphics[width=\textwidth]{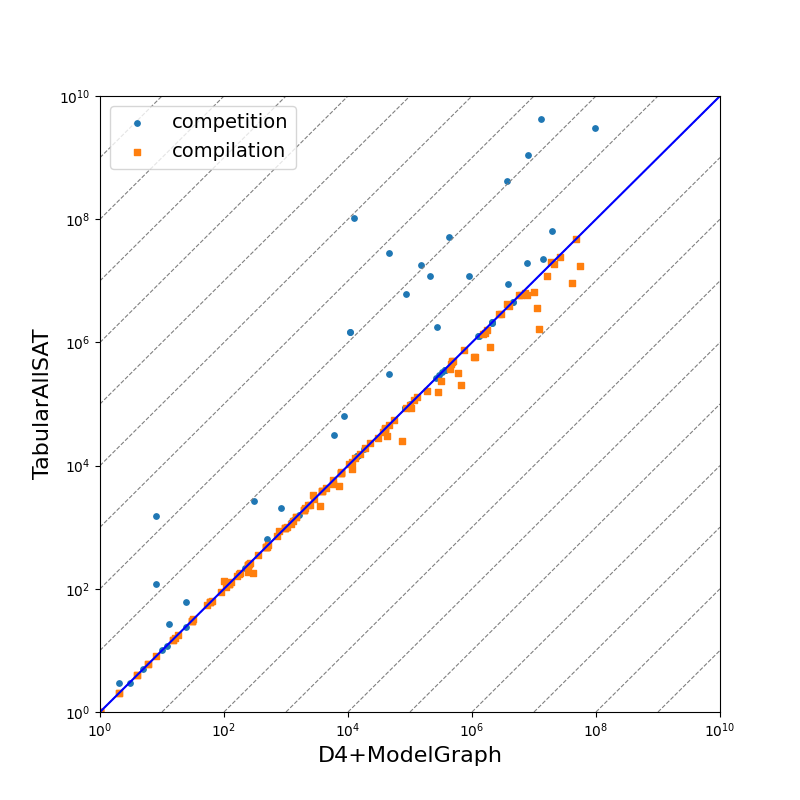}
        \caption{\# of assignments}    
        \label{fig:soa-hard-ass}
    \end{subfigure}
    \caption{Scatter plots comparing \solverPlus{} against {\sc D4+ModelGraph} on AllSAT problems in \cite{lagniez2024leveraging}. The $x$ and $y$ axes are log-scaled.} 
    \label{fig:stateofart-hard}
    \scriptsize
    \begin{tabular}{ccc}
                    & \solverPlus{} & {\sc D4+Modelgraph}    \\
        compilation(197) & 137 & {\bf 165} \\ 
        competition(246) & 99 & {\bf 113} \\ \hline
        Total(443) & 236 & {\bf 289}
    \end{tabular}
    \caption{Number of harder instances solved by \solverPlus{} and {\sc D4+ModelGraph} within a timeout (1200 seconds) for the AllSAT benchmark.}
    \label{tb:table-timeout3}

\end{figure}

In these experiments, we considered BC, NBC, and BDD \cite{toda2016implementing}, respectively a blocking, a non-blocking, and a BDD-based disjoint AllSAT solver. BC also provides the option to obtain partial assignments (from now on \textsc{BC\_Partial}). We also considered \textsc{MathSAT5} \cite{cimatti2013mathsat5}, since it provides an interface to compute partial enumeration of propositional problems by exploiting blocking clauses, and the very-recent enumeration approach {\sc D4+ModelGraph} from \cite{lagniez2024leveraging} that enumerates formulas after transforming them into an equivalent d-DNNF representation.
Other AllSAT solvers, such as {\sc BASolver} \cite{zhang2020accelerating} and {\sc AllSATCC} \cite{liang2022allsatcc}, are currently not publicly available, as reported also by other papers \cite{fried2023allsat}.

We evaluated the computational performance of \solverPlus{} against several state-of-the-art solvers using the same benchmark set we used in \cite{spallitta2024disjoint}. The primary objective of this evaluation was to demonstrate that the new chronological implicant shrinking algorithm in \solverPlus{} does not degrade performance in AllSAT problems compared to the previous version. Figure \ref{fig:stateofart} presents scatter plots comparing \solverPlus{} with other state-of-the-art solvers. The results align with those reported in \cite{spallitta2024disjoint}, showing that \solverPlus{} performs competitively or even much better than almost all solvers. The only exception is against the two AllSAT algorithms based on knowledge compilation, respectively {\sc BDD} and {\sc D4+ModelGraph}. Both approaches perform better than \solverPlus{} when the problem instances contain few clauses (which is the case of {\it rnd3sat} problems); in this case, the knowledge compilation procedure is less resource-intensive. 

For the sake of completeness, we opted for a more extensive evaluation of \solverPlus{} and {\sc D4+ModelGraph} by using the benchmarks proposed in \cite{lagniez2024leveraging}. In this case, with no surprise, \solverPlus{} is outperformed by the approach based on knowledge compilation, in alignment with results in \cite{lagniez2024leveraging}. We must remark, however, several points. First, the new datasets are based on model counting competition and used for knowledge compilation testing, thus they heavily rely on pre-processing techniques such as partitioning or AND-gate decomposition.  In addition to that, and as also stated in \cite{lagniez2024leveraging}, \solverPlus{} has a lower memory footprint, never experiencing timeouts during execution, making our tool better suited for situations where memory resources are limited. Finally, it is worth noting that for most of the tested problems, the tools returned the same number of models (see the bisector line in Figure \ref{fig:soa-hard-ass}). In most instances, the model count is equivalent to the number of partial assignments retrieved. This indicates that the structure of these problems prevents implicant shrinking from eliminating even a single atom, thus limiting our algorithm's ability to demonstrate its full potential.

\section{From AllSAT to Projected AllSAT}
\label{sec:projected}

\subsection{Algorithms and implementation}

To extend Alg. \ref{algo:chronocdcl} for projected enumeration, we consider a formula $F$ with two mutually exclusive sets of variables: relevant variables $V_r$ and irrelevant variables $V_i$. Recall that for a formula $F(V_r, V_i)$, an assignment $\mu_r$ projected over $V_r$ satisfies $F$ if $\mu_r$ satisfies $\exists V_i.F$. The core search algorithm itself remains unchanged: when we generate a total assignment $\eta$ during our search loop, we can partition it into $\eta_r$ and $\eta_i$, corresponding to the relevant and irrelevant variables, respectively. Thus, $\eta_r$ represents the assignment that the algorithm should ultimately produce. As a result, the conflict analysis component of the algorithm does not require modifications. However, it is crucial to note that we are working within the framework of disjoint enumeration. Therefore, the chronological implicant shrinking procedure must be adapted to prevent repetitions while effectively pruning irrelevant variables from the total assignment.

A fundamental adjustment is needed in the variable ordering heuristic, where we prioritize relevant variables over irrelevant ones. This ensures that once the last relevant variable is assigned, any subsequent decision literal from irrelevant variables can be safely ignored. Recalling \sref{sec:ordering}, if a non-relevant decision literal $\ell$ is assigned before the relevant ones, the blocking clause subsumed by chronological backtracking would include $\ell$, preventing its removal to guarantee disjointness. By prioritizing relevant variables during the decision phase, we ensure that every partial assignment satisfies $F$ without introducing non-relevant literals as decision literals.

We now discuss how the implicant shrinking algorithm is influenced by projection. All modifications needed for the projected enumeration extensions are highlighted in \textcolor{darkgreen}{green}. Specifically, when determining which literals to drop from $T$, we can skip all literals corresponding to variables in $V_i$, as these irrelevant variables would be dropped regardless. Thus, in Alg. \ref{alg:lift-aggressive}, line \ref{alg:safe} we ensure the procedure skips non-relevant literals and lifts them anyway. 
It is important to note that unit propagation in Alg. \ref{alg:reorder}, line \ref{alg:unitit} might force some non-relevant literals back into the trail. However, this does not affect the correctness of the procedure, since no non-relevant variable can be a decision variable this way. Moreover, only literals corresponding to relevant variables are included in the partial model when the model is printed. Algorithm \ref{algo:get-assignment-full}, line \ref{algo:print} is updated accordingly, ensuring only variables in $V_r$ are considered.

\subsection{Experimental evaluation}
\label{sec:exp-projected}

All the additional ideas discussed in \sref{sec:projected} to integrate projection in the algorithm have been added in \solverPlus{}. Experiments are performed on an Intel Xeon Gold 6238R @ 2.20GHz 28 Core machine with 128 GB of RAM, running Ubuntu Linux 22.04. Timeout has been set to 1200 seconds.

\subsubsection{Comparison of implicant shrinking algorithms}

 \begin{figure*}[t!]
        \centering
        \begin{subfigure}[b]{0.32\textwidth}
            \centering
            \includegraphics[width=\textwidth]{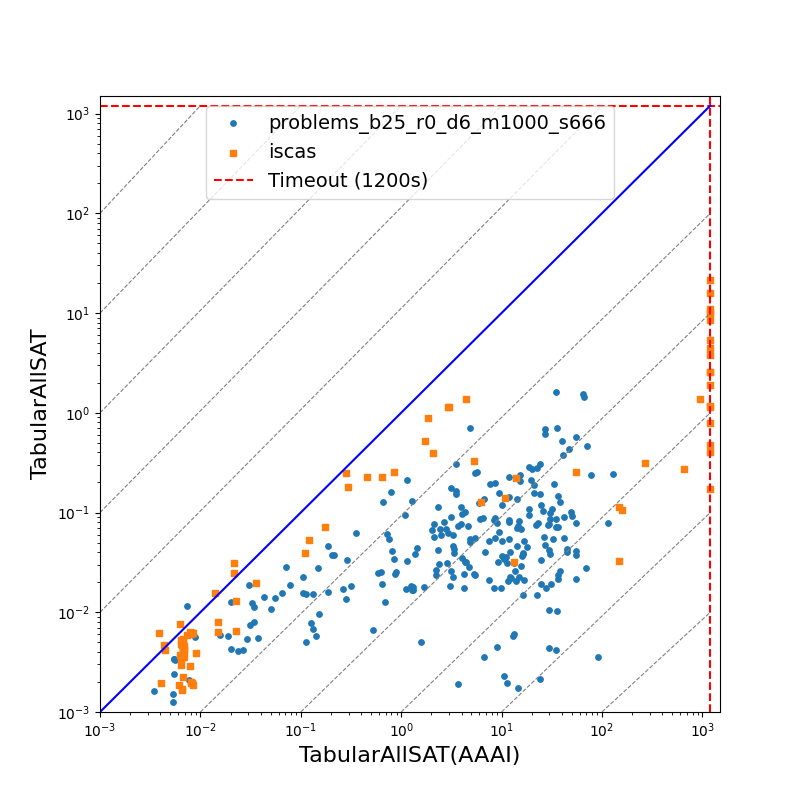}
            \caption%
            {{Time (in seconds)}}    
            \label{fig:proj-abl-time}
        \end{subfigure}
        \begin{subfigure}[b]{0.32\textwidth}  
            \centering 
            \includegraphics[width=\textwidth]{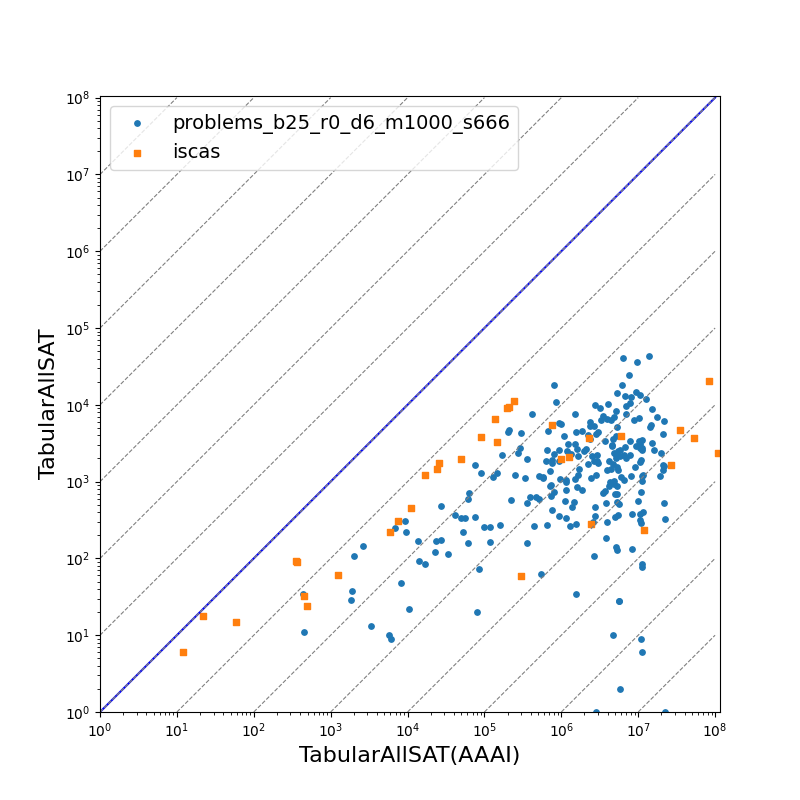}
            \caption[]%
            {{\# of partial assignments}}    
            \label{fig:proj-abl-ass}
        \end{subfigure}
        \caption{Scatter plot comparing CPU time and number of partial assignments generated of \solverPlus{} against the implicant shrinking algorithm of \solver{} on projected AllSAT problems. The $x$ and $y$ axes are log-scaled.} 
        \label{fig:scatter-projected-ablation}
    \end{figure*}

We started by comparing the two implicant shrinking algorithms from \sref{sec:aggressive}, to ensure that the new method does not introduce negative side effects in projected enumeration, and thus how the more aggressive pruning of total assignments is beneficial for projected enumeration. To evaluate \solverPlus{} on projected AllSAT enumeration, we focused on non-CNF instances that require preprocessing into CNF before conversion to the DIMACS format. This preprocessing step introduces additional CNF-specific variables irrelevant to the final enumeration task. Therefore, we selected benchmarks inspired by \cite{masina2023cnf}, specifically: ($i$) 250 synthetic benchmark instances containing non-CNF formulas with double implications, each with 25 Boolean variables and a formula depth of 6, and ($ii$) a set of 100 instances from the $iscas$ benchmark suite. All of these problems were originally generated as non-CNF formulas. For the CNF transformation, we chose the approach proposed in \cite{masina2023cnf}, as it is more suitable for enumeration and the generation of compact partial assignments than the classic Plaisted-Greenbaum transformation.

The results, shown in Figure \ref{fig:scatter-projected-ablation}, indicate that the new implicant shrinking approach positively impacts projected enumeration, yielding improvements in both execution time and the number of partial assignments generated.

\subsubsection{Comparison against state-of-the-art solvers}

\begin{figure}[t]
    \centering
    \begin{subfigure}[b]{0.32\textwidth}
        \centering
        \includegraphics[width=\textwidth]{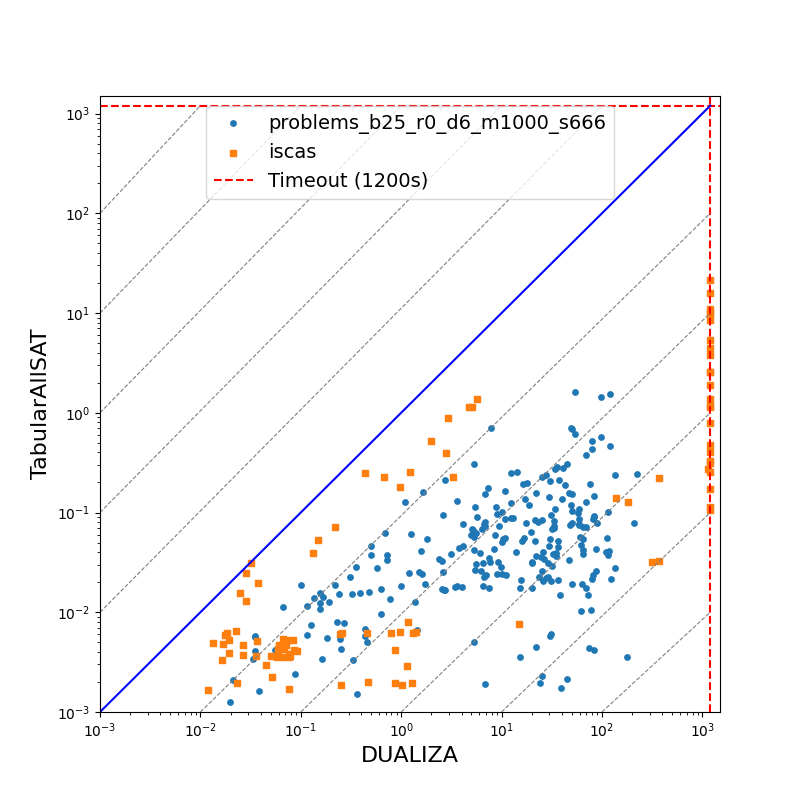}
        \caption{{\textsc{Dualiza}}}    
        \label{fig:proj-dualiza}
    \end{subfigure}
    \begin{subfigure}[b]{0.32\textwidth}  
        \centering 
        \includegraphics[width=\textwidth]{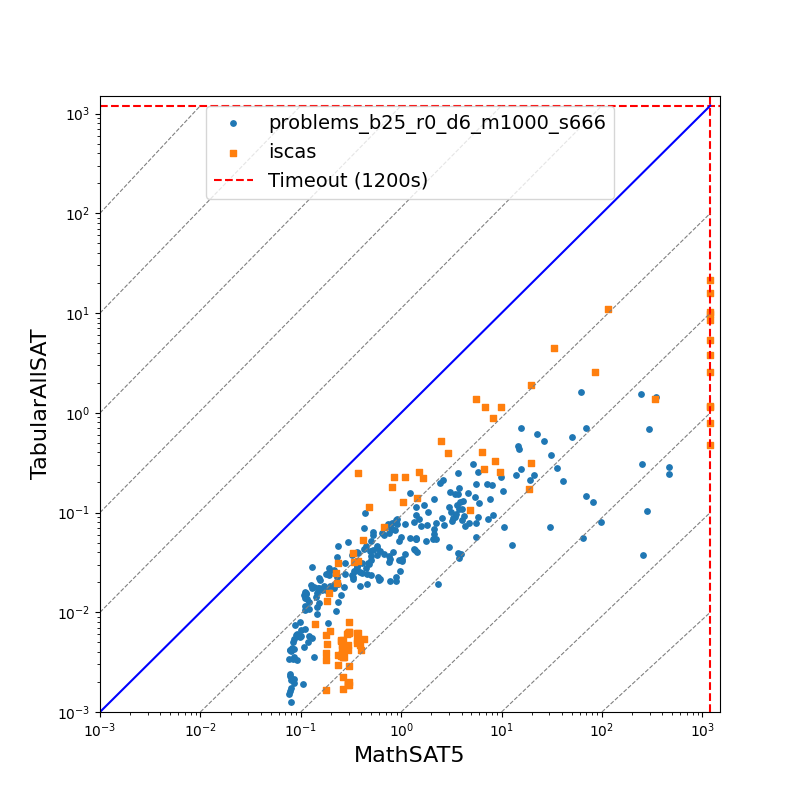}
        \caption{{\textsc{MathSAT5}}}    
        \label{fig:proj-mathsat}
    \end{subfigure}
     \begin{subfigure}[b]{0.32\textwidth}  
        \centering 
        \includegraphics[width=\textwidth]{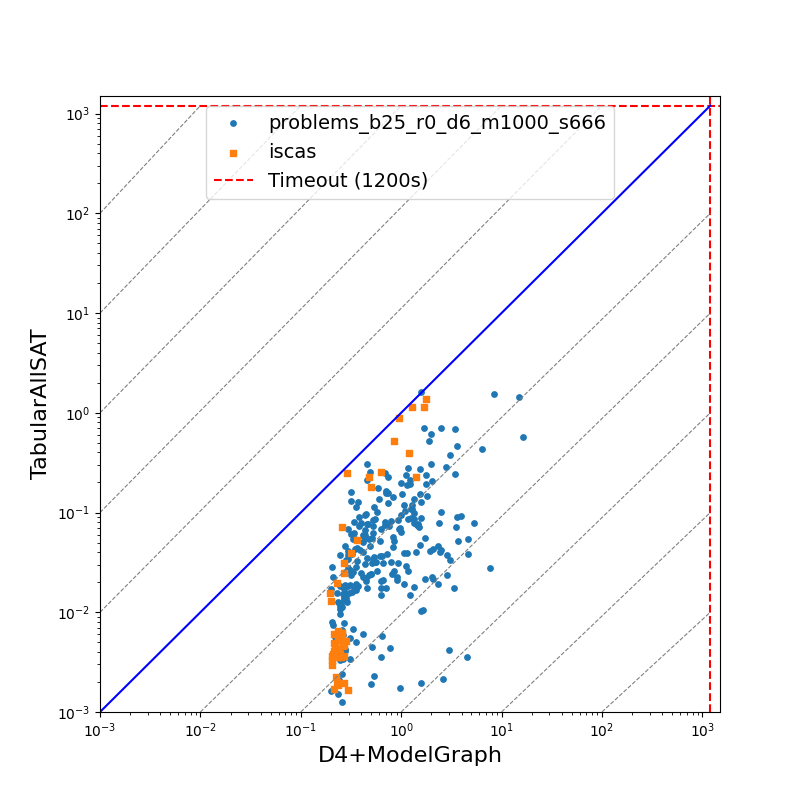}
        \caption{{\textsc{D4+ModelGraph}}}    
        \label{fig:proj-d4}
    \end{subfigure}

    \caption{Scatter plot comparing CPU time of \solverPlus{} against \textsc{Dualiza}, \textsc{MathSAT5}, and \textsc{D4+ModelGraph} on projected AllSAT problems. The $x$ and $y$ axes are log-scaled.} 
    \label{fig:scatter-projected}

    \vspace{0.5cm}

    \scriptsize
    \begin{tabular}{ccccc}
                  & \solverPlus{} & \textsc{D4+ModelGraph} & \textsc{Dualiza} & \textsc{MathSAT5} \\
    iscas (100)       & \textbf{100} & \textbf{100}   & 76    & 78    \\
    b25\_r0\_d6 (250) & \textbf{250} & \textbf{250}   & \textbf{250}   & 174   \\ \hline
    Total (350)       & \textbf{350} & \textbf{350}   & 326   & 252  
    \end{tabular}
    \caption{Table reporting the number of instances solved by each solver within the timeout time (1200 seconds) for projected AllSAT benchmark.}
    \label{tb:timeout-proj}    
\end{figure}

We compared \solverPlus{} against ($i$) {\sc MathSAT5}, ($ii$) {\sc Dualiza}, a model counter and AllSAT solver that utilizes dual reasoning \cite{mohle2018dualizing}, and ($iii$) the {\sc D4+ModelGraph} tool from \cite{lagniez2024leveraging}. We remark that, despite \cite{lagniez2024leveraging} not directly addressing projected enumeration, it is possible to use {\sc D4} to generate projected d-DNNF, on top of which {\sc ModelGraph} can retrieve partial assignments. The results, depicted in Figure \ref{fig:scatter-projected}, demonstrate that \solverPlus{} significantly outperforms both {\sc Dualiza} and {\sc MathSAT5}, generating partial assignments much faster. 
\ignore{In the case of {\sc MathSAT5}, this performance difference may be partly due to {\sc MathSAT5}'s primary role as an SMT solver, meaning its AllSAT functionality may not be as optimized as needed for this specific task. It is also worth noting, however, the limited availability of publicly accessible projected AllSAT solvers in the literature, which makes {\sc MathSAT5} one of the few tools that can be used for experimental evaluation and highlights the relevance of these findings.} 

\begin{figure*}[t!]
        \centering
        \begin{subfigure}[b]{0.32\textwidth}
            \centering
            \includegraphics[width=\textwidth]{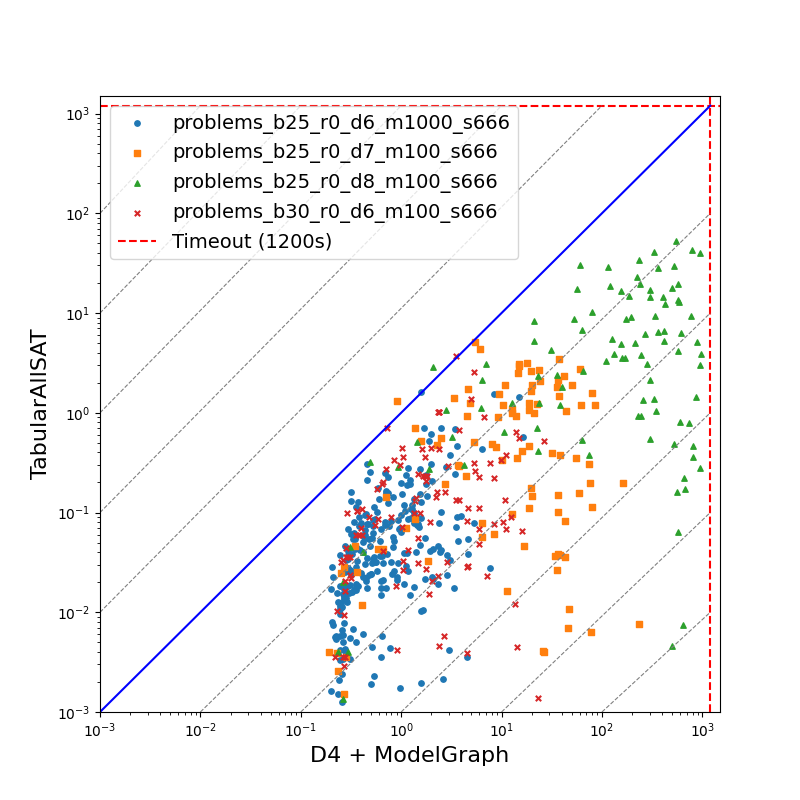}
            \caption%
            {{Time}}    
            \label{fig:proj-time-hard}
        \end{subfigure}
        \begin{subfigure}[b]{0.32\textwidth}  
            \centering 
            \includegraphics[width=\textwidth]{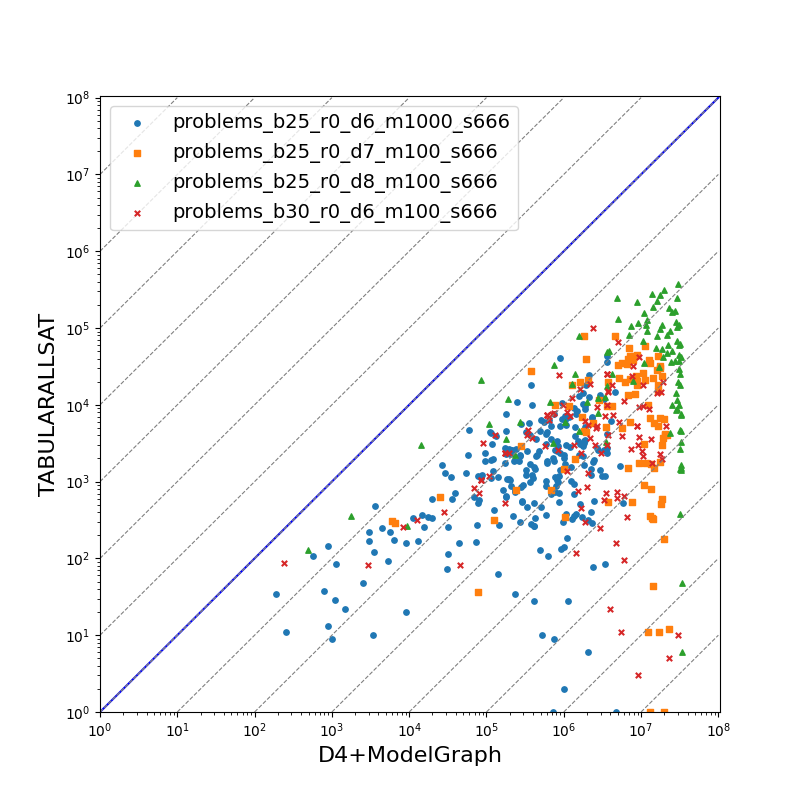}
            \caption[]%
            {{\# of partial assignments}}    
            \label{fig:proj-ass-hard}
        \end{subfigure}
        \caption{Scatter plot comparing CPU time and number of partial assignments of \solverPlus{} against \textsc{D4+ModelGraph} on harder projected AllSAT problems. Notice that in this batch of experiments no timeout happens. The $x$ and $y$ axes are log-scaled.} 
        \label{fig:scatter-projected-hard}
    \end{figure*}

We notice that the performance gap against {\sc D4+ModelGraph} is not as dramatic as with respect to the other two solvers, so for the sake of completeness, we tested both tools using more complex synthetic benchmarks, where {\sc MathSAT5} and {\sc Dualiza} reached timeout for almost every file. We generated 2 additional benchmarks with 25 Boolean variables and formula depth 7 and 8, respectively. We also provided a set of benchmarks with depth 6 and 30 variables, to check if adding more important variables does impact enumeration. The results, shown in Figure \ref{fig:scatter-projected-hard}, now clearly show the superiority of \solverPlus{} against the knowledge compilation approach, both considering computation times and number of partial assignments retrieved.

\section{From AllSAT to AllSMT}
\label{sec:SMT}

\subsection{Algorithms and implementation}

To extend the algorithm to address first-order logic theories, the search algorithm should integrate a theory solver and call it to check if the current assignment that satisfies the Boolean abstraction of a formula \(F\) is also theory consistent. These additional checks affect the definition of some of the algorithms of \solver{}, and all changes to the original \solver{} algorithms are colored in \textcolor{blue}{\textbf{blue}}.

First, in Alg. \ref{algo:chronocdcl} once a total trail has been generated, we must verify if there are theory inconsistencies. We perform a \(T\)-consistency check (Alg. \ref{algo:chronocdcl}, line \ref{alg:tcheck}) and, if a \(T\)-conflict is generated, then we must analyze the conflict and backjump accordingly (Alg. \ref{algo:chronocdcl}, lines \ref{alg:tcheck-bad}-\ref{alg:tcheck-bad-end}).

Second, once a literal has been decided and \textsc{UnitPropagation} is executed, there could be some other \(T\)-atoms that are implied by the newly added literals in \(T\) or there could be a \(T\)-conflict. For this reason, \textsc{UnitPropagation} now is a two-step procedure: (i) propositional unit propagation, to satisfy the Boolean abstraction of \(F\), and (ii) \(T\)-propagation, which also works as an early pruning algorithm in SMT solving. If either unit propagation call generates a conflict, \textsc{AnalyzeConflict} is executed. The \(T\)-conflict analysis does not differ from the Boolean conflict analysis algorithm, so additional changes are not required in Alg. \ref{algo:conflict}. We must remark, however, that theory solvers can add new \(T\)-atoms during execution, e.g., in \(\mathcal{L}\mathcal{I}\mathcal{A}\), the branch-and-bound algorithm could generate new inequalities atoms. If this happens, then the trail maximum size increases, and all new \(T\)-atoms are flagged as non-relevant variables.

\subsection{Experimental Evaluation}
\label{sec:exp-allsmt}

We implemented all the ideas discussed in the paper into \solverSMT{}, whose executable file and all benchmarks are available on Zenodo \cite{spallitta_2024_bis}. An updated version of the source code is available at \url{https://github.com/giuspek/tabularAllSMT.git}. \solverSMT{} integrates {\sc MathSAT5} as the theory solver, which is under a proprietary license, thus \solverSMT{} code is not publicly available, but the executable file is provided. Experiments are performed on an Intel Xeon Gold 6238R @ 2.20GHz 28 Core machine with 128 GB of RAM, running Ubuntu Linux 22.04. Timeout has been set to 1200 seconds.

\begin{figure}[t]
    \centering
    \begin{subfigure}[b]{0.32\textwidth}  
        \centering 
        \includegraphics[width=\textwidth]{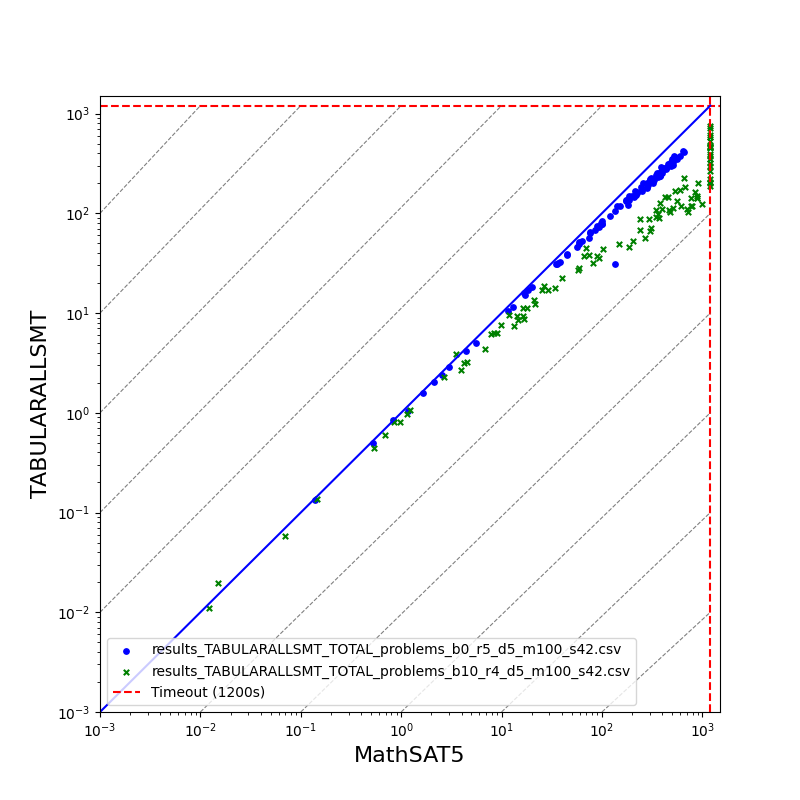}
        \caption{Time - Total enumeration}  
        \label{fig:smt-total}
    \end{subfigure}
    \begin{subfigure}[b]{0.32\textwidth}
        \centering
        \includegraphics[width=\textwidth]{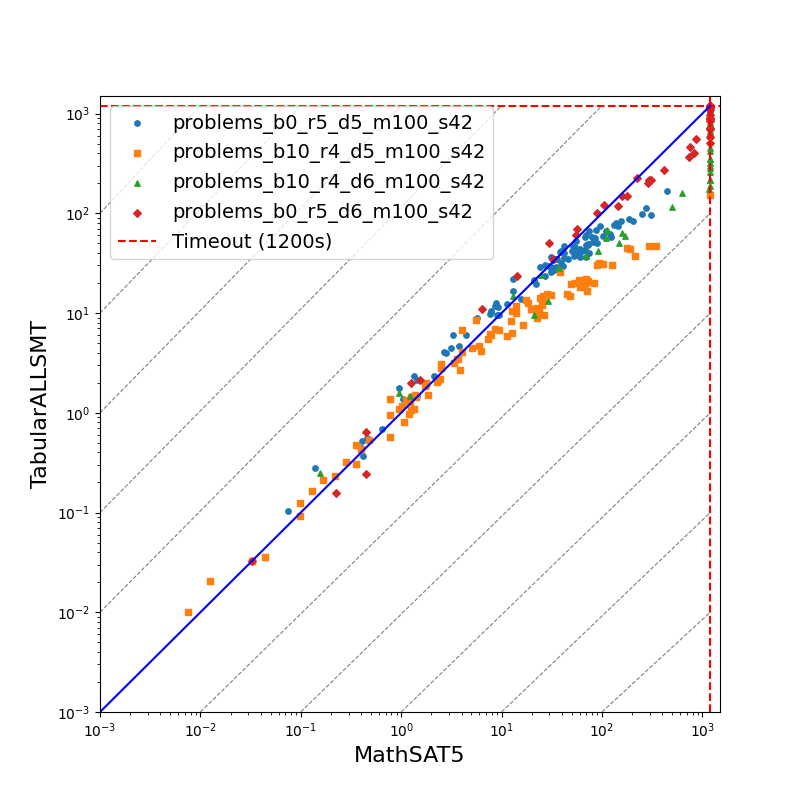}
        \caption{Time - Partial enumeration}    
        \label{fig:smt-partial}
    \end{subfigure}
    \begin{subfigure}[b]{0.32\textwidth}  
        \centering 
        \includegraphics[width=\textwidth]{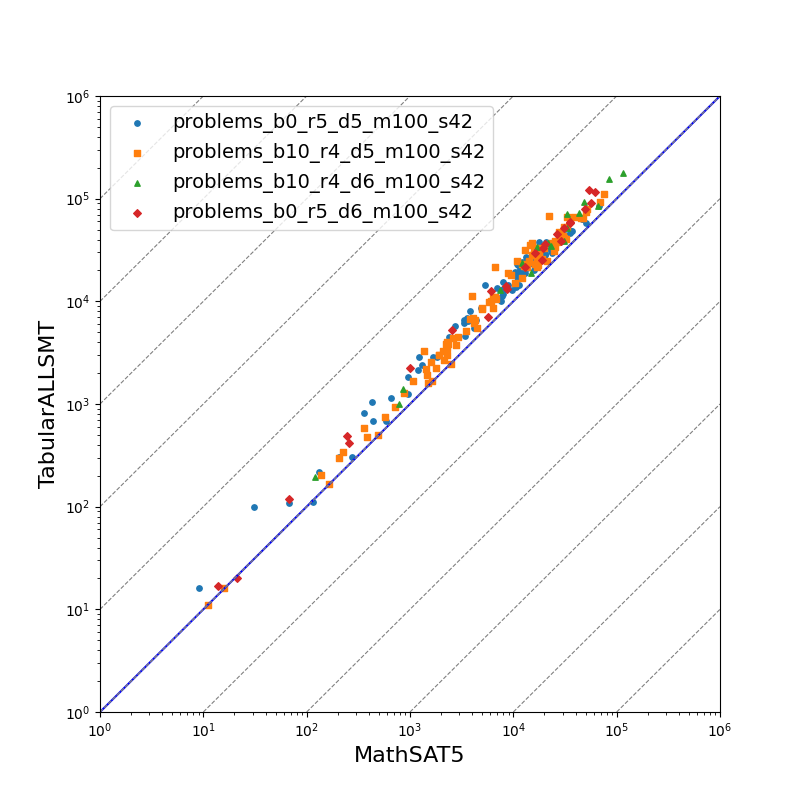}
        \caption{\# of partial assignments}    
        \label{fig:smt-nass}
    \end{subfigure}

    \caption{Scatter plot comparing CPU time for total enumeration (a), CPU time for partial enumeration (b), and the number of partial assignments for partial enumeration (c) generated by \solverSMT{} against {\sc MathSAT5} on AllSMT problems. The $x$ and $y$ axes are log-scaled.} 
    \label{fig:allsmt}

    \vspace{0.5cm}

    \small
    \begin{tabular}{ccc}
                  & \solverSMT{} &  \textsc{MathSAT5} \\
    b0\_r5\_d5 (100)       & \textbf{100}   &  \textbf{100}    \\
    b10\_r4\_d5 (100) & \textbf{100}    & 99   \\ 
    b0\_r5\_d6 (100)       & \textbf{48}     & 18    \\
    b25\_r0\_d6 (100) & \textbf{49}    & 26   \\ \hline
    Total (400)       & \textbf{287}   & 243  
    \end{tabular}
    \caption{Table reporting the number of instances solved by each solver within the timeout time (1200 seconds) for AllSMT benchmark.}
    \label{tb:timeout-smt} 

\end{figure}

For the final set of experiments, we used benchmarks inspired by \cite{masina2024cnf}, generating several synthetic benchmarks with varying numbers of Boolean variables ($b$), real variables ($r$), and formula depth ($d$). We compared \solverSMT{} against {\sc MathSAT5}, which is currently the only publicly available projected AllSMT solver.

We began by evaluating the effect of blocking clauses during total SMT enumeration. This experiment was designed to demonstrate how the introduction of blocking clauses as in {\sc MathSAT5} negatively impacts performance compared to our algorithm, particularly when dealing with first-order logic theories. We performed total enumeration on two smaller benchmark sets, and the results, presented in the scatter plot in Figure \ref{fig:smt-total}, clearly illustrate that several instances are successfully solved by \solverSMT{}, whereas {\sc MathSAT5} reaches the timeout limit. Since the theory reasoning is shared among the two tools, it is evident that the lack of blocking clauses is the primary factor contributing to these timeouts.

We continue by comparing results on disjoint partial enumeration, including the same benchmarks used for Figure \ref{fig:smt-total} plus two other datasets with higher depth. The results, shown in Figure \ref{fig:smt-partial} and \ref{fig:smt-nass}, indicate that while {\sc MathSAT5} is slightly more effective at shrinking assignments into shorter partial models, \solverSMT{} outperforms it over the long term considering CPU times, especially as the complexity of instances increases. This difference is primarily due to the exponential number of blocking clauses that {\sc MathSAT5} adds, which eventually hampers its performance. Additionally, Table \ref{tb:timeout-smt} presents the number of problems solved within the timeout limits, further emphasizing that \solverSMT{} successfully solves a significant number of problems that {\sc MathSAT5} cannot handle. It is important to highlight that early pruning, implemented in both {\sc MathSAT5} and \solverSMT{}, can significantly reduce the Boolean search space and, as a result, the number of calls to the T-solver. However, while early pruning can be beneficial, it may also lead to unnecessary calls to the T-solver. In the context of enumeration, this overhead can negatively affect performances, as the extra solver calls introduce additional computational cost, explaining why results are less impactful than those shown in \sref{sec:projected}.

We remark that the experiments in this section focus on linear real arithmetic; \solverSMT{}, however, is compatible with all theories accepted by {\sc MathSAT5}.

\end{gschange}

\section{Conclusion}
\label{sec:conclusion}

In this work, we introduced \solverPlus{} and \solverSMT{}, two new solvers designed for efficient projected enumeration in AllSAT and AllSMT, respectively. By combining CDCL and chronological backtracking, we addressed the inherent inefficiencies of traditional blocking solvers, avoiding the performance degradation caused by excessive blocking clauses. Our novel aggressive implicant shrinking algorithm further reduced the number of partial assignments with respect to its predecessor, ensuring a more compact representation of the solution space. We extended our solver framework to support projected enumeration and SMT formulas, integrating theory reasoning into the search process. Extensive experimental results showed that our solvers outperform existing state-of-the-art techniques, offering a significant advantage in both propositional and SMT-based problems.



\FloatBarrier

\bibliographystyle{elsarticle-num} 
\bibliography{aaai24}


\end{document}